\documentclass[aps, prd, twocolumn, amsmath, floats,floatfix, superscriptaddress, nofootinbib]{revtex4-2}

\usepackage{blindtext}
\usepackage{afterpage}
\usepackage[export]{adjustbox}
\usepackage{wrapfig}
\usepackage{natbib}
\usepackage{graphicx,color}
\usepackage{amsmath,amssymb}
\usepackage{verbatim}
\usepackage{float}
\usepackage{wasysym}
\usepackage{amssymb,subfigure}
\usepackage{epsfig}
\usepackage{psfrag}
\usepackage{dsfont}
\usepackage{amsfonts}
\usepackage{mathrsfs}
\usepackage{multirow}
\usepackage{soul}
\usepackage{times}
\usepackage{bm}
\usepackage{hyperref}
\usepackage{xspace}
\usepackage[normalem]{ulem}
\usepackage[export]{adjustbox}
\usepackage{orcidlink}
\usepackage{mathtools}
\usepackage{cleveref}
\hypersetup{
  colorlinks=true,        
  linkcolor=blue,         
  citecolor=blue,         
  }
\usepackage{pifont}
\usepackage[normalem]{ulem}

\usepackage{multirow}
\usepackage{verbatim}

\graphicspath{{./PLOTS/}}                         
\newcommand{\beq}{\begin{equation}} 
\newcommand{\eeq}{\end{equation}} 
\newcommand{\beqn}{\begin{eqnarray}} 
\newcommand{\eeqn}{\end{eqnarray}}

\newcommand{\zD}{{\raise1.0ex\hbox{${}^{\ \circ}$}}\!\!\!\!\!D}
\newcommand{\alone}{{\raise0.5ex\hbox{${}^{\ 1}$}}\!\!\!\!\alpha}

\newcommand{\nalam}{\mathrel{\raise0.9ex\hbox{$^\lambda$}\mkern-14mu
\lower0.0ex\hbox{$\nabla$}}}

\newcommand{\zeroD}{{\raise1.0ex\hbox{${}^{\ \circ}$}}\!\!\!\!\!D}

\newcommand{\zLap}{{\raise1.0ex\hbox{${}^{\ \circ}$}}\!\!\!\!\Delta}
\newcommand{\zna}{{\raise1.0ex\hbox{${}^{\ \circ}$}}\!\!\!\!\!\nabla}
\newcommand{\zS}{{\raise1.0ex\hbox{${}^{\ \circ}$}}\!\!\!\!\!S}

\newcommand{\Ms}{M_\odot}
\newcommand{\be}{\begin{equation}}
\newcommand{\ee}{\end{equation}}

\def\QEQ{{%
    \setbox0\hbox{$I$}%
    \rlap{\hbox to \wd0{\hss--\hss}}\box0
}}

\begin{document}

\title{On the treatment of thermal effects in the equation of state on neutron star merger remnants}                                                                                                   
\author{Davide Guerra\orcidlink{0000-0003-0029-5390}}\thanks{E-mail: davide.guerra@uv.es}
\affiliation{Departamento de Astronom\'{\i}a y Astrof\'{\i}sica, Universitat de Val\'encia,
  Dr. Moliner 50, 46100, Burjassot (Val\`encia), Spain}
\author{Milton Ruiz\orcidlink{0000-0002-7532-4144}}
\affiliation{Departamento de Astronom\'{\i}a y Astrof\'{\i}sica, Universitat de Val\'encia,
  Dr. Moliner 50, 46100, Burjassot (Val\`encia), Spain}
\author{Michele Pasquali\orcidlink{0000-0002-5784-0700}}
\affiliation{Department of Chemistry, Life Sciences and Environmental Sustainability, University of Parma, 43124 Parma, Italy}                                                                                                   
\author{Pablo Cerd\'a-Dur\'an\orcidlink{0000-0003-4293-340X}}
\affiliation{Departamento de Astronom\'{\i}a y Astrof\'{\i}sica, Universitat de Val\'encia,
  Dr. Moliner 50, 46100, Burjassot (Val\`encia), Spain}
\affiliation{Observatori Astron\'omic, Universitat de Val\'encia,
C/ Catedr\'atico Jos\'e Beltr\'an 2, 46980, Paterna (Val\'encia), Spain}
\author{Arnau Rios\orcidlink{0000-0002-8759-3202}}
\affiliation{
 Departament de F\'isica Qu\`antica i Astrof\'isica, 
Universitat de Barcelona, Mart\'i i Franqu\`es 1, E08028 Barcelona, Spain
}
\affiliation{
 Institut de Ci\`encies del Cosmos (ICCUB),
Universitat de Barcelona, Mart\'i i Franqu\`es 1, E08028 Barcelona, Spain
}

\author{Jos\'e A. Font\orcidlink{0000-0001-6650-2634}}
\affiliation{Departamento de Astronom\'{\i}a y Astrof\'{\i}sica, Universitat de Val\'encia,
  Dr. Moliner 50, 46100, Burjassot (Val\`encia), Spain}
\affiliation{Observatori Astron\'omic, Universitat de Val\'encia,
C/ Catedr\'atico Jos\'e Beltr\'an 2, 46980, Paterna (Val\'encia), Spain}
\date{\today}


\begin{abstract}
We present results from long-term, numerical-relativity simulations of binary neutron star (BNS) mergers modeled using both, fully tabulated, finite-temperature, equations of state (EOSs) and their corresponding {\it hybrid} representations (combining a cold piecewise-polytropic part and a thermal ideal-gas-like part). The simulations extend up to $\gtrsim 150 \,\rm ms$ which allows us to assess the role of the treatment of finite-temperature effects on the dynamics of the hypermassive neutron star (HMNS) remnant. Our study focuses on the analysis of the spectra of the post-merger gravitational-wave (GW) signals and on how these are affected by the treatment of thermal effects in the two EOS representations. This allows us to derive new quasi-universal relations for the main post-merger GW frequencies, associated with oscillation modes of the remnant, and examine the impact of thermal effects thereon. Our simulations highlight distinct differences in the GW frequency evolution related to the thermal modeling of the EOS, demonstrating that deviations from established quasi-universal relations become significant at late post-merger phases. Furthermore, we investigate the stability of the HMNS against convection. Employing both the Ledoux criterion, necessary condition for the development of convective instabilities, and the Solberg-Høiland criterion, a generalized criterion for axisymmetric perturbations based on a combined analysis of the Brunt-V\"ais\"al\"a frequency and of the epicyclic frequency, we show that differential rotation and thermal stratification in the HMNS give rise to local (yet sustained) convective patterns that persist beyond $\sim 100\,\rm ms$ after merger. Those convective patterns, while substantially different between tabulated and hybrid EOS treatments, trigger the the excitation of inertial modes with frequencies smaller than those attained by the fundamental quadrupolar mode, and are potentially within reach of third-generation GW detectors. The late-time excitation of inertial modes, previously reported in studies based on hybrid EOS, is fully supported by the tabulated, finite-temperature EOS simulations presented here, which account for thermal effects in a more consistent way.   
\end{abstract}

\maketitle

\section{Introduction}

The detection of the first gravitational-wave (GW) signal from the merger of a binary neutron star (BNS), {\tt GW170817}, is a major milestone of GW astronomy~\cite{Abbott:2017qsu}. This multi-messenger observation~\cite{Drout:2017ijr,LIGOScientific:2017zic} has established a direct connection between GW sources and electromagnetic counterparts. The analysis of the inspiral part of the GW signal has impacted many fields of research providing, among others, (i) tight constraints on the equation of state (EOS) of neutron stars (NSs) at supranuclear densities~\cite{gw170817,Radice2018,Ruiz:2017due,Rezzolla:2017aly,Shibata:2017xdx}, (ii) an independent measure for the expansion of the Universe~\cite{LIGOScientific:2017adf,Dietrich:2020efo}, and (iii) evidence of ejecta masses of  $\sim 0.01-0.05M_\odot$ with velocities~$\sim 0.1-0.3\,c$~\cite{Cowperthwaite:2017dyu,Kasliwal:2017ngb,Smartt:2017fuw}.  This ejecta is roughly consistent with the estimated r-process production rate required to explain the Milky Way r-process abundances. Complementary insights into the internal structure of NSs are expected to be obtained through the observation 
of the post-merger GW signal, not yet fully accessible to the sensitivity of present-day detectors at high frequency. In particular, the post-merger signal may allow to construct quasi-universal (EOS-insensitive) relations between the main oscillation-mode frequencies and physical properties of the NS,  such as its tidal deformability or its radius~\cite{Topolski:2023ojc,Takami2015,bauswein2015exploring}, potentially constraining masses and radii of NSs as well as the EOS at finite temperature. 

The fate of the BNS merger remnant is primarily influenced by the system's total mass and the EOS. Depending on the differential rotation and the maximum mass supported by the EOS, the merger can result in either a \textit{supramassive neutron star} (SMNS) that survives the collision and produces a stable NS or eventually collapses into a black hole (BH) over secular timescales, or a \textit{hypermassive neutron star} (HMNS) that collapses to a BH on dynamical timescales~\cite{Sarin_2021}. On the other hand, massive enough remnants may promptly form a BH right after merger.  Understanding the role of the EOS is crucial for interpreting observational data, as different EOS impact the evolution of the remnant in different ways, also affecting the GW emission. Therefore, two main approaches of EOS are used in numerical simulations of BNS mergers. The first approach are {\it hybrid} EOSs, which combine a zero-temperature component with an ideal-gas-like thermal pressure term to account for thermal energy in the post-merger remnant. The cold section is typically approximated using a \textit{piecewise polytropic} representation~\cite{Read2009}. This approach, by far the most commonly employed, enables efficient simulations by approximating thermal effects, facilitating the exploration of variations in the EOS stiffness and thermal behavior across a large density range~\cite{Endrizzi:2016kkf,Ozel:2016oaf,Maione_2016,Feo2016Modeling,DePietri:2018tpx,DePietri2020,Figura:2020fkj,Palenzuela:2021gdo}. On the other hand, BNS merger simulations also use, albeit to a lesser extent, {\it tabulated} EOS. These EOSs are derived from nuclear physics models and include dependencies on temperature, composition, and density, providing high accuracy in modeling NS matter. Tabulated EOSs are particularly valuable for capturing thermal effects, which are essential in post-merger dynamics where shock heating and high temperatures significantly influence the remnant's behavior~\cite{PhysRevLett.108.011101,Fields_2023}.

In this work we present a suite of new long-term, numerical-relativity simulations of BNS mergers with a focus on the treatment of temperature effects of the EOS on the fate of the post-merger remnant. The initial data are chosen such that most of the simulations (all models but one) lead to the formation of a long-lived HMNS which survives for timescales longer than $100\,\rm ms$ after merger. Our study is motivated by the work of~\cite{DePietri:2018tpx,DePietri2020} who conducted such long-term BNS merger simulations using hybrid EOSs. Complementary long-term efforts have adopted a hybrid General Relativistic Magnetohydrodynamics (GRMHD) strategy that couples a full-GR evolution during merger to a conformally-flat (CFC) treatment for the secular remnant, enabling post-merger evolutions out to $\sim1\,\rm s$ while reducing computational costs by factors, thus opening the way to systematic secular studies of BNS remnants~\cite{Ng:2023yyg,Armengol:2021mbt}. In our case, we improve the consistency of the treatment of thermal effects by accounting for tabulated, finite-temperature EOS. This goes beyond standard thermal treatments in terms of polytropes or parametrizations, like those presented in Refs.~\cite{Raithel:2023zml,Raithel:2023gct,Raithel:2025geu}. Some of this previous work indicates that thermal effects have a small influence into the GW properties of BNS. We find that a fully different treatment of thermal effects (from hybrid to tabulated) can indeed have a much more significant effect on the remnant formation and evolution.
Hereafter, we refer to those two sets of models as {\tt tab} and {\tt hyb}, respectively. For each of the two approaches we simulate four EOSs, namely {\tt SLy4}, {\tt DD2}, {\tt HShen}, and {\tt LS220}, to explore a range of stiffness and thermal properties, examining how these variations influence the dynamics from the early to the late post-merger phases. \\

Following~\cite{DePietri:2018tpx,DePietri2020,Shibata:2025nfj}, we study the (local) convective stability analysis in the bulk of the HMNS at sufficiently long timescales up to $140\,\rm ms$ after merger, by employing a generalized stability criterion based of a combined analysis of the Brunt-V\"ais\"al\"a frequency and of the epyciclic frequency, the so-called \textit{Solberg-Høiland criterion} for axisymmetric convective stability~\cite{Tassoul:2000} as well as the general relativistic version of the Ledoux criterion. Convection triggers the excitation of inertial modes in the remnant, with the Coriolis force serving as a primary restoring force. Inertial modes have frequencies lower than that of the fundamemtal quadrupolar $f$-mode, dominant in the early post-merger, and contribute to GW emission at reduced amplitudes. Therefore, we examine local convective patterns in our simulations for each of our models. \\
Moreover, by analyzing the excitation of oscillation modes, especially the late-time inertial modes, we investigate the influence of different EOS representations on the thermal and rotational evolution of the remnant. Our approach allows us to assess how each EOS affects the long-term stability of the remnant and the characteristics of the GW signal that third-generation detectors might observe~\cite{2017CQGra..34d4001A,Punturo2010ET}. For further details on detectability prospects using the simulations reported here, see Refs.~\cite{VillaOrtega2024,Miravet2024}.\\

\begin{figure}
\begin{center}
\includegraphics[width=1.\columnwidth]{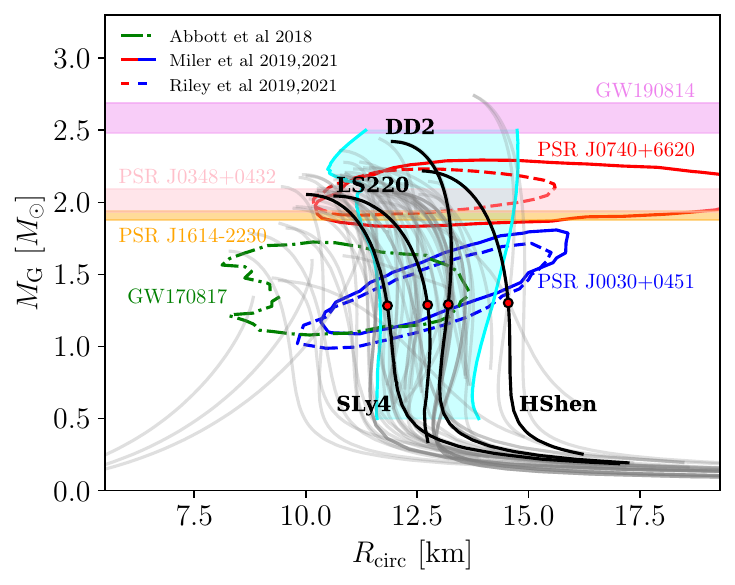}
\caption{Gravitational mass vs circumferential radius for several EOSs along with some current observational constraints:
the mass measurements of two high-mass pulsars and the observational constraints (within $95\%$ confidence levels) 
from NICER/XMM-Newton and LIGO-Virgo, along with the constraints from nuclear physics experiments, modeled through a Typel-Wolter density-dependent relativistic mean-field functional
 (within $99\%$ confidence levels) from~\cite{prasanta} (cyan region). The black curves represent the thermal, 
 tabulated EOSs employed in this work, while the gray curves correspond to cold EOSs~\cite{doi:10.1146,compose}. 
 The red points indicate the specific models used in our study.}
\label{fig:constrains}
\end{center}
\end{figure}

This paper is organized as follows: In Section~\ref{sec:initial} we detail our initial data setup and numerical methods, explaining the EOS models and how their different representations affect the equilibrium models. In Section~\ref{sec:results_early} we examine the early post-merger phase, emphasizing the role of the dominant quadrupolar $f$-mode and its effects on the GW spectrum, considering alternative parametrizations of the standard $7-$pieces and $10-$pieces  polytropic models for the hybrid representation. In Section~\ref{sec:latepostmerger} we address the intermediate and late post-merger phases, highlighting the presence of convective instabilities and the significance of inertial modes and their sustained excitation for next-generation GW observatories. We also discuss the importance of the definition of temperature for BNS merger simulations and the role temperature plays in the tabulated and hybrid approaches. Lastly, in this section we also examine the characteristics of the $f$-mode and inertial modes and their correlation with the HMNS rotational profiles. Our conclusions are summarized in Section~\ref{Conclusions}. The paper ends with Appendix~\ref{app:Temperature} where further technical aspects on the definition of temperature are provided. Unless indicated otherwise, we employ units in which $c = G = \Ms = 1$.

\begin{figure*}
\begin{center}
\hspace{-5mm}
\includegraphics[width=2.12\columnwidth]{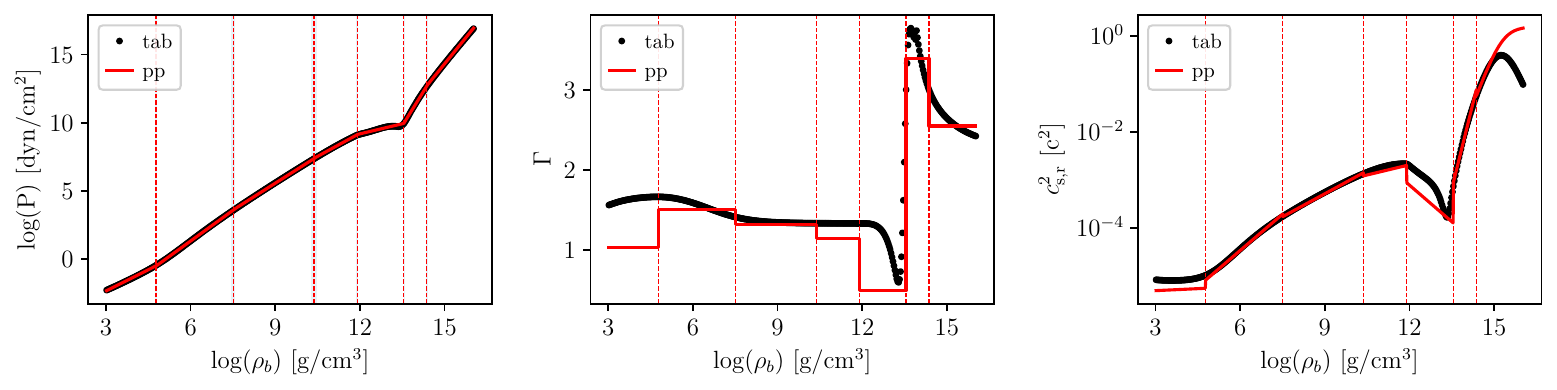}
\caption{From left to right the panels depict the pressure $P$ in log scale, adiabatic index $\Gamma$ and relativistic sound speed $c_\mathrm{s,r}^2$ against the barion density (in cgs units), for the {\tt LS220} EOS at the lowest available temperature of $T=0.01$ MeV in $\beta$-equilibrium. Black dots indicate tabulated points while red lines show the piecewise polytropic representation of the EOS. Vertical dashed lines mark the six breakpoints obtained from the regression code~\cite{Pilgrim2021}.}
\label{fig:PvsRho}
\end{center}
\end{figure*}

\section{Initial data and methods}
\label{sec:initial}
\subsection{Choosing and building EOSs}
\label{subsec:eoss}

We select four finite-temperature, fully tabulated microphysical EOSs, namely {\tt SLy4}~\cite{Chabanat97un}, {\tt DD2}~\cite{Typel:2009sy}, {\tt HShen}~\cite{Shen:2011qu}, and {\tt LS220}~\cite{Lattimer91}. The EOS tables are publicly available at~\cite{stellarcollapse,Schneider:2017tfi}. The relation between the gravitational mass and the circumferential radius for each NS model built with these EOSs is plotted in Fig.~\ref{fig:constrains} (black solid curves). The selected models span a wide range of physical properties, including radii between $\sim10\,\rm km$ to $\sim16\,\rm km$ and maximum gravitational masses (for the corresponding spherical NS) between $\sim2M_\odot$ to $\sim2.5M_\odot$. We note that the lowest publicly available temperature for these EOSs is $T = 0.01\,\rm{MeV}$. Correspondingly, the many gray curves in Fig.~\ref{fig:constrains} correspond to cold EOSs, and have been obtained from~\cite{doi:10.1146,compose}. Some of these models consider quarks, pion and kaon condensates and hyperons. \\

Fig.~\ref{fig:constrains} also displays current observational and theroretical constraints on the gravitational mass and circumferential radius of NSs. In particular, the plot includes the observational constraints on PSRJ0030+0451~\cite{Riley_2019,Miller_2019}, PSRJ0740+6620~\cite{Riley:2021pdl,Miller_2021}, GW170817~\cite{gw170817}, the masses for NS systems  PSRJ0348+0432~\cite{psrj0348} and PSRJ1614-2230~\cite{psrj1614}, and the constraint for the mass of the secondary component of compact binary system GW190814~\cite{Abbott_2020}. Confidence intervals at $95\%$ $(2\sigma)$ are indicated for every constraint, calculated using publicly available posteriors from the previous references. We also plot predictions from the unconstrained Typel-Wolter (TW) density-dependent relativistic mean-field (RMF) model~\cite{prasanta}. This shows the mass-radius relation derived from the ensemble of TW-based EOS sampled using only nuclear physics-informed priors—ranges of nuclear matter parameters constrained by nuclear models and experimental data, but not yet filtered by astrophysical constraints. The TW prior is displayed in cyan lines as a $99\%$ confidence interval contour, representing the model prediction spread for NS mass-radius relations based solely on these priors. Finally, the red points in Fig.~\ref{fig:constrains} indicate the specific configurations we choose for our simulations, with the rest mass fixed at $M_0 = 1.4\,M_\odot$. We opt to fix this value rather than the gravitational mass $M_G$ because the latter can vary during the construction of both single star and binary star systems.

This study aims to assess the impact of the EOS treatment on the BNS merger and post-merger dynamics and GW emission. 
A complete tabulated NS EOS includes essential thermodynamic quantities as functions of baryon density $\rho$, temperature $T$, and electron fraction $Y_e$, defined as the ratio of electron number density $n_e$ to total baryon number density $n_B$, $Y_e = n_e / n_B$. This ratio also reflects the proton-to-baryon ratio, since charge neutrality indicates that $Y_e$ equals the proton fraction. In tables derived from the \textsc{StellarCollapse} repository~\cite{stellarcollapse}, this holds true; however, tables from the \textsc{CompOSE} repository~\cite{compose} may report the charge fraction $Y_q = Y_e + Y_\mu$, where $Y_\mu$ is the muon fraction. (See also~\cite{Werneck:2022exo,Etienne:2015cea} for further information on the use of tabulated EOS.)
To construct a NS in equilibrium, the EOS must be cold, solely depending on baryon density $\rho$. We achieve this by selecting the lowest available temperature, $T = T_{\rm min} = 0.01 \, \rm MeV$, from the full-tabulated EOS. Additionally, we impose $\beta-$equilibrium, ensuring that the rates of weak interaction processes ($\beta$ decay and electron capture) are balanced. This equilibrium occurs when the chemical potentials satisfy the relation,
\begin{equation}
    \mu_n = \mu_p + \mu_e + \mu_\nu\ ,
\end{equation}
where $\mu_n$, $\mu_p$, $\mu_e$, and $\mu_\nu$ are the neutron, proton, electron, and neutrino chemical potentials, respectively. In a cold NS, the absence of trapped neutrinos results in $\mu_\nu = 0$, meaning $\beta-$equilibrium depends only on the neutron, proton, and electron chemical potentials. Hence, for the given temperature $T = T_{\rm min}$, we adjust $Y_e$ to satisfy $\Delta \mu = \mu_n - (\mu_p + \mu_e) = 0$, producing the cold-tabulated EOS in $\beta-$equilibrium.

Correspondingly, in the hybrid EOS representation all thermodynamic quantities $Q$ are expressed as 
\begin{equation}
    Q(\rho,T) = Q_{\rm cold}(\rho) + Q_{\rm th}(\rho,T)\ .
\end{equation}
When $Q$ is the pressure, the cold component uses a piecewise polytropic EOS while the thermal component employs a simple polytropic equation with a fixed thermal index, such that
\begin{equation}
    \label{eq:pthermal}
    P_{\rm th}(\rho,T) = \kappa_{\rm th}\ \rho ^ {\Gamma_{\rm th}}\ .
\end{equation}
In this work, the thermal index is set to $\Gamma_{\rm th}=1.8$. 

\noindent A simple linear regression code~\cite{Pilgrim2021} is used to derive the piecewise polytropic representation ($7$-pieces) of the lowest temperature table at $\beta$-equilibrium in terms of $Q_{\rm cold}(\rho)$. This code takes the pressure-density relationship as input to determine the optimal piecewise polytropes that most accurately align with the original data from the table, as illustrated in Fig.~\ref{fig:PvsRho} for the case of {\tt LS220}. The breakpoints are selected automatically in order to obtain a $\chi^2$ value close to $1$. The approach varies from the one described in~\cite{Read:2008iy} in which the seven pieces are divided in two parts: four for the low-density area based on the {\tt SLy} EOS, and three parts for the high-density region obtained by fitting the tabulated data. In our setup, the location of breakpoints differ for each EOS but the number of breakpoints is constant. Our choice of number of breakpoints follows previous studies of BNS merger simulations based on $7$-piece piecewise polytropic EOSs (e.g.~\cite{Hotokezaka:2011dh,Carney2018Comparing,De_Pietri_2016,Feo2016Modeling}).

In addition, in Fig.~\ref{fig:PvsRho} we illustrate how the hybrid EOS representation approximates other quantities in terms of the barion density $\rho_b$. For the pressure, the red solid line closely matches the tabulated data represented by the black dots; however, the differences are more pronounced for the adiabatic index $\Gamma \equiv \partial\log P / \partial\log\rho_b$.  These derivatives are computed numerically for tabulated data, while the piecewise polytropic approximation assumes a constant adiabatic index for each segment, derived from the regression code~\cite{Pilgrim2021}. Similar differences are found for the sound speed, defined as,
\begin{equation}
    c_\mathrm{s,r}^2 = \left . \frac{\partial P}{\partial \epsilon} \right\rvert_{s} =  \Gamma_i \frac{P}{\epsilon + P}\ .
\end{equation}
Here, $c_\mathrm{s,r}$ represents the relativistic definition of the sound speed, with $\epsilon$ being the total energy density and the derivatives are calculated at constant entropy $s$. As for $\Gamma$, we use numerical derivatives for the tabulated EOS, while for the piecewise polytropic approximation, the equations transform to the form at the right side of the equation, since $\epsilon = \rho_b + P/(\Gamma_i - 1)$, where $\Gamma_i$ denotes the adiabatic index for the $i-$piece of the piecewise representation. Discrepancies between the numerical sound speed and the analytical approximation are evident from Fig.~\ref{fig:PvsRho}. The most pronounced differences are found in the barion density range $\rho_b \in [10^{10}, 10^{15}]\, \rm g/cm^3$, for all EOSs.

\subsection{Constructing single star equilibrium configurations}
\label{subsec:massradius}

\begin{figure}
\begin{center}
\includegraphics[width=0.9\columnwidth]{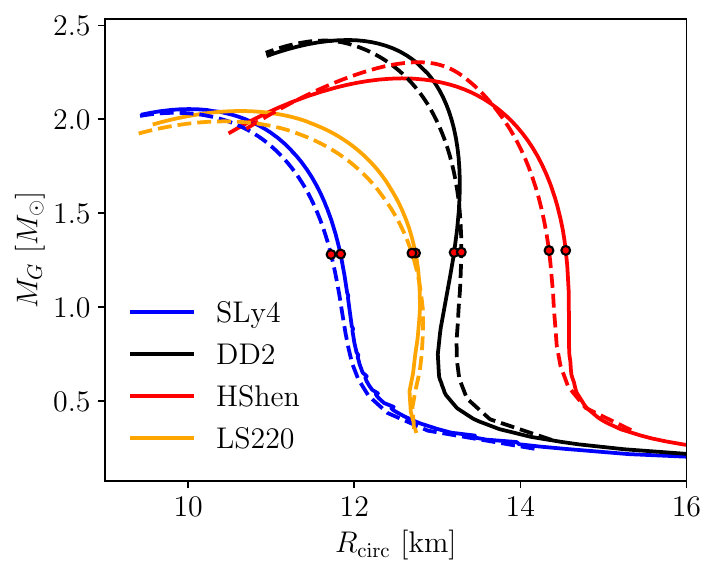}
\caption{Gravitational mass versus circumferential radius for spherically symmetric NSs built with all EOSs of our sample. Solid lines represent tabulated EOS configurations and dashed lines piecewise polytropic approximations. Red dots mark NS configurations with a baryon mass of $1.4\,\Ms$.}
\label{fig:MvsRinitial}
\end{center}
\end{figure}

Before discussing initial data for BNS mergers, we first examine equilibrium configurations of (cold) isolated spherically-symmetric stars. This analysis is conducted with the two distinct approaches for the EOS discussed in the previous section. We investigate this aspect to assess the impact on the initial data of both, the EOS representation and the number of pieces of the hybrid representation, to try and make the data as similar as possible and minimize potential initial discrepancies. 

The equilibrium configurations were built with the \textsc{LORENE} code~\cite{Lorene} employing a grid consisting of 33 radial points and 17 angular points (equally distributed in $\theta$). The results, plotted in Fig.~\ref{fig:MvsRinitial} and reported in Table~\ref{table:singleNS} for the $7-$pieces polytropic EOS, show slight variations between the two EOS representations, predominantly seen in the radii of the stellar models. The analysis revealed that {\tt HShen} showed the greatest discrepancy in radius between the original table and the piecewise polytropic representation, followed by {\tt SLy4} and {\tt DD2}. In contrast, models built with the {\tt LS220} EOS showed the least difference between the two approaches, except for the central baryon density value $\rho_{\rm c}$. 

\begin{figure*}
\begin{center}
\includegraphics[width=2.\columnwidth]{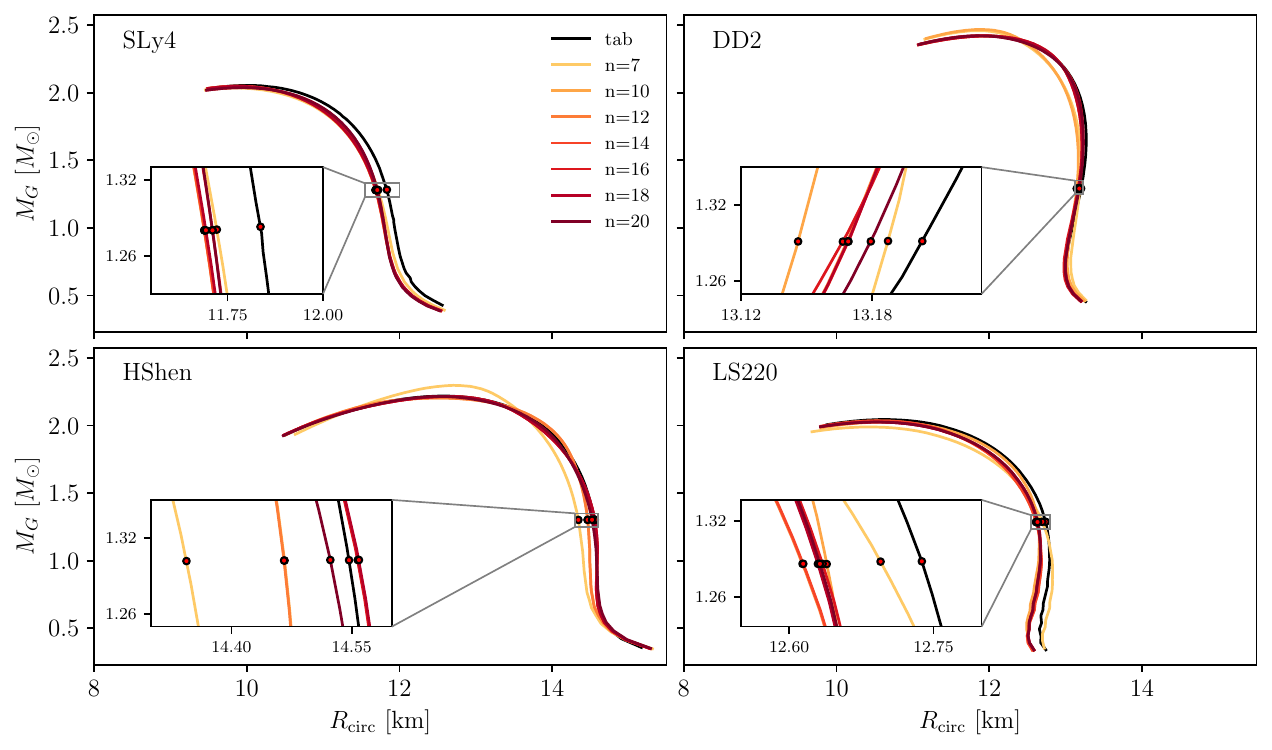}
\caption{Dependence on the number of pieces of the gravitational mass versus circumferential radius curves for spherically symmetric NSs for all EOSs. In each panel, solid colored lines indicate the number of pieces of the piecewise polytropic representation while the solid black line represents the tabulated version of the corresponding EOS. The red dots indicate  configurations with baryon mass $M_0=1.4 \Ms$.}
\label{fig:diff_pwp}
\end{center}
\end{figure*}

\begin{center}
\begin{table}[th]
\caption{Summary of the initial properties of the single NS configurations.
  We list the EOS, the gravitational mass $M_G[\Ms]$, the equatorial radius $R_{\rm eq}[\rm km]$, the circumferential radius $R_{\rm circ}[\rm km]$, the compactness $\mathcal{C}=M_G/R_{\rm circ}$, and the central baryon density $\rho_{\rm c} [\rm g/cm^3]$. All configurations have a fixed baryon mass $M_0 = 1.4\,\Ms$. The tag for each configuration is composed of the EOS name followed by suffix `tab' or `hyb(7)' indicating if a tabulated or a $7-$pieces piecewise polytropic EOS has been used to build the stellar model. The last four rows of the table report the relative errors in percentage for each quantity.
\label{table:singleNS}}
\begin{tabular}{cccccc}
\multicolumn{1}{l}{} & \multicolumn{1}{l}{} & \multicolumn{1}{l}{} & \multicolumn{1}{l}{} & \multicolumn{1}{l}{} & \multicolumn{1}{l}{} \\ \hline
EOS      & $M_G$    & $R_{\rm eq}$& $R_{\rm circ}$ & $\mathcal{C}$ & $\rho_{\rm c}\times10^{14}$ \\ \hline
{\tt SLy4$_{\rm tab}$}  & 1.28     & 9.85     & 11.84    & 0.160    & 8.23  \\
{\tt DD2$_{\rm tab}$}   & 1.29     & 11.21    & 13.20    & 0.144    & 5.52  \\
{\tt HShen$_{\rm tab}$} & 1.30     & 12.55    & 14.55    & 0.132    & 4.69  \\
{\tt LS220$_{\rm tab}$} & 1.29     & 10.75    & 12.74    & 0.149    & 6.70  \\ \hline
{\tt SLy4$_{\rm hyb(7)}$}  & 1.28     & 9.74     & 11.72    & 0.161    & 8.48  \\
{\tt DD2$_{\rm hyb(7)}$}   & 1.29     & 11.30    & 13.29    & 0.144    & 5.70  \\
{\tt HShen$_{\rm hyb(7)}$} & 1.30     & 12.35    & 14.34    & 0.134    & 4.89  \\
{\tt LS220$_{\rm hyb(7)}$} & 1.29     & 10.71    & 12.69    & 0.150    & 6.98  \\ \hline \hline
{\tt SLy4$_{\rm err}$}  & 0.17\%   & 1.16\%   & 0.98\%   & 0.81\%   & 3.06\%    \\
{\tt DD2$_{\rm err}$}   & 0.11\%   & 0.76\%   & 0.66\%   & 0.55\%   & 3.27\%    \\
{\tt HShen$_{\rm err}$} & 0.06\%   & 1.61\%   & 1.39\%   & 1.34\%   & 4.31\%    \\
{\tt LS220$_{\rm err}$} & 0.02\%   & 0.41\%   & 0.35\%   & 0.33\%   & 4.16\%    \\ \hline
\multicolumn{1}{l}{} & \multicolumn{1}{l}{} & \multicolumn{1}{l}{} & \multicolumn{1}{l}{} & \multicolumn{1}{l}{} & \multicolumn{1}{l}{}
\end{tabular}
\end{table}
\end{center}

Next, we constructed piecewise polytropic EOSs varying the number of breakpoints in baryon density. We increased the number of pieces from $7$ to $25$ pieces to check whether this increase yields results closer to the original tabulated versions for each EOS. Fig.~\ref{fig:diff_pwp} demonstrates that adding more breakpoints does not necessarily produce equilibrium configurations closer to the tabulated case. While convergence is observed for the {\tt HShen} case, this is not true for other EOSs. For configurations with $M_{\rm G} = 1.4\,\Ms$, the {\tt SLy4} EOS does not show consistent changes in the mass-radius curves. In contrast, the {\tt DD2} and {\tt LS220} cases exhibit a larger error in the radius $R_{\rm circ}$, reaching  $\sim 100/150$ meters. This uncertainty in the radius measurement, along with the compactness and central baryon densities, is similar to the error associated with the numerical resolution of the grid we use for evolving BNS mergers (see below where we compare results for the $7$-piece and the $10$-piece hybrid version of each EOS).

We also examine the effect on the initial data of varying the low-density cut in configurations with $7$ and $10$ pieces. Following~\cite{Pilgrim2021}, we generated the two piecewise representations from a tabulated version with a minimum density of $\rho_{\rm min}\sim10^3\,{\rm g/cm^3}$ and another one with $\rho_{\rm min}\sim10^4\,{\rm g/cm^3}$. This changes the breakpoints to optimize the piecewise polytropic fit, achieving a $\chi^2\sim 1$. We find that the results are highly dependent on the choice of the EOS: for the {\tt SLy4} case, the low-density cut has minimal impact, while it significantly affects other cases. 

The previous analysis highlights the inherent difficulty of building fully consistent, identical initial data when using two different EOS representations, even for the simplest possible case of cold, spherically symmetric stellar models. This issue automatically translates to the case of BNS initial data, by the way such initial data is built. In order to assess the magnitude of the differences, a comparison of BNS merger evolutions using tabulated and hybrid EOSs derived from $7-$pieces and $10-$pieces polytropic representations will be discussed in Section~\ref{sec:results_early}.

\subsection{Initial data and setup for the BNS simulations}
\label{subsec:IDbinary}

\begin{center}
  \begin{table}[th]
    \caption{Summary of the initial properties of the BNS configurations.
      The columns report the EOS, the gravitational mass $M_G[\Ms]$, the compactness $\mathcal{C}$, the maximum stable mass under linear perturbations $M^{\rm max}[\Ms]$, the tidal deformability $\Lambda=(2/3)\kappa_2\,\mathcal{C}^{-5}$ (for each individual star; $\kappa_2$ is the second Love number), the ADM mass $M_{\rm ADM}[\Ms]$, the ADM angular momentum $J_{\rm ADM}[\Ms^2]$ and the angular velocity $\Omega [\rm krad/s]$ of the binary. The tag for each configuration includes the EOS name labeled by `tab' or `hyb' indicationg the type of EOS representation used for the initial data. The number in parenthesis next to `hyb' specifies the number of pieces used for the piecewise polytropic approximation.
      \label{table:Iparameters}}
        \begin{tabular}{ccccccccc}
          \hline
          \hline
           EOS & $M_G$      & $\mathcal{C}$    & $M^{\rm max}$ &  $\Lambda$   & $M_{\rm ADM}$  & $J_{\rm ADM}$ & $\Omega$\\
          \hline
            {\tt SLy4$_{\rm tab}$}       & 1.28   & 0.161   & 2.055   & 536.00   & 2.54          & 6.63          & 1.77 \\
            {\tt DD2$_{\rm tab}$}        & 1.29   & 0.144   & 2.422   & 1098.68  & 2.56          & 6.73          & 1.78 \\
            {\tt HShen$_{\rm tab}$}      & 1.30   & 0.132   & 2.218   & 1804.67  & 2.58          & 6.83          & 1.78 \\
            {\tt LS220$_{\rm tab}$}      & 1.29   & 0.149   & 2.044   & 851.72   & 2.55          & 6.68          & 1.77 \\
           \hline
           \hline
            {\tt SLy4$_{\rm hyb(7)}$}       & 1.28   & 0.161   & 2.035   & 511.71   & 2.54          & 6.62          & 1.77 \\
            {\tt DD2$_{\rm hyb(7)}$}        & 1.29   & 0.144   & 2.419   & 1113.92  & 2.56          & 6.73          & 1.78 \\
            {\tt HShen$_{\rm hyb(7)}$}      & 1.30   & 0.134   & 2.304   & 1633.24  & 2.58          & 6.82          & 1.78 \\
            {\tt LS220$_{\rm hyb(7)}$}      & 1.29   & 0.150   & 1.989   & 899.05   & 2.55          & 6.69          & 1.77 \\
           \hline
           \hline
            {\tt SLy4$_{\rm hyb(10)}$}      & 1.28   & 0.162   & 2.046   & 507.49   & 2.54          & 6.62          & 1.77 \\
            {\tt DD2$_{\rm hyb(10)}$}       & 1.29   & 0.145   & 2.468   & 1063.44  & 2.56          & 6.72          & 1.78 \\
            {\tt HShen$_{\rm hyb(10)}$}     & 1.30   & 0.133   & 2.204   & 1734.38  & 2.58          & 6.82          & 1.78 \\
            {\tt LS220$_{\rm hyb(10)}$}     & 1.29   & 0.150   & 2.025   & 915.47   & 2.55          & 6.69          & 1.78 \\
           \hline\hline
        \end{tabular}
  \end{table}
\end{center}

To build initial data for irrotational BNS systems we consider equal-mass binaries in quasi-equilibrium circular orbits in the late inspiral phase. These initial configurations are generated using the same code used for constructing the single-star equilibrium configurations in the previous section, \textsc{LORENE}~\cite{Gourgoulhon:2000nn}. The binaries consist of two identical irrotational NSs modeled by the four finite-temperature (fully tabulated) microphysical EOSs of our sample.\\ 

The properties of our BNS models are summarized in Table~\ref{table:Iparameters}. Each NS has a rest mass of $M_0 = 1.4\,\Ms$, which is a typical mass observed in binary pulsar systems~\cite{Lattimer:2012nd}. The ADM quantities reported in the table are computed from the initial data and represent the total mass-energy and angular momentum of the system as measured at spatial infinity. The inspection of the maximum gravitational mass ($M^{\rm max}$) across different EOS shows that increasing the number of pieces in polytropic approximations from $7$ to $10$ does not consistently enhance the accuracy relative to the tabulated EOSs. For {\tt SLy4} and {\tt HShen}, the $10-$pieces approximation offers a slight improvement over the $7-$pieces version. However, for {\tt DD2}, the $7-$pieces approximation is actually closer to the tabulated value than the $10-$pieces approximation. The discrepancies are minimal, typically within a few percent, suggesting that additional pieces offer negligible improvement to the approximation.\\

The numerical evolution of the initial data was performed using the \textsc{IllinoisGRMHD} code~\cite{Werneck:2022exo,Etienne:2015cea,Noble:2005gf}, which is a component of the broader \textsc{Einstein Toolkit} infrastructure~\cite{EinsteinToolkit:2024_05,Loffler:2011ay}. The \textsc{IllinoisGRMHD} code evolves the spacetime fields in the Baumgarte-Shapiro-Shibata-Nakamura (BSSN) formulation of the Einstein equations~\cite{Baumgarte:1998te,Shibata:1995we}, coupled with the puncture gauge conditions with fourth-order spatial differentiation. The damping coefficient in the shift condition is set to $1/M_{\rm ADM}$. This choice helps to prevent coordinate singularities and allows for long-term stable  evolutions of the spacetime fields~\cite{Alcubierre:2002kk}. Additionally, the \textsc{IllinoisGRMHD} code employs the Valencia formulation for the general relativistic hydrodynamics (GRHD) equations~\cite{Banyuls:1997,Font:2008fka}, involving a high-resolution shock-capturing (HRSC) finite-volume algorithm for the integration of the matter fields. For time integration, the method of lines is used with a fourth-order Runge-Kutta scheme. The Courant-Friedrichs-Lewy (CFL) factor is fixed at $0.5$ to ensure stability and accuracy in the simulations. Further details on the numerical infrastructure and methodology can be found in Refs.~\cite{Werneck:2022exo,Etienne:2015cea,Noble:2005gf}, providing in-depth information for interested readers. Specifically, Ref.~\cite{Noble:2005gf} discusses the numerical techniques employed in magnetohydrodynamic simulations.\\

\begin{figure*}
\begin{center}
\includegraphics[width=2.\columnwidth]{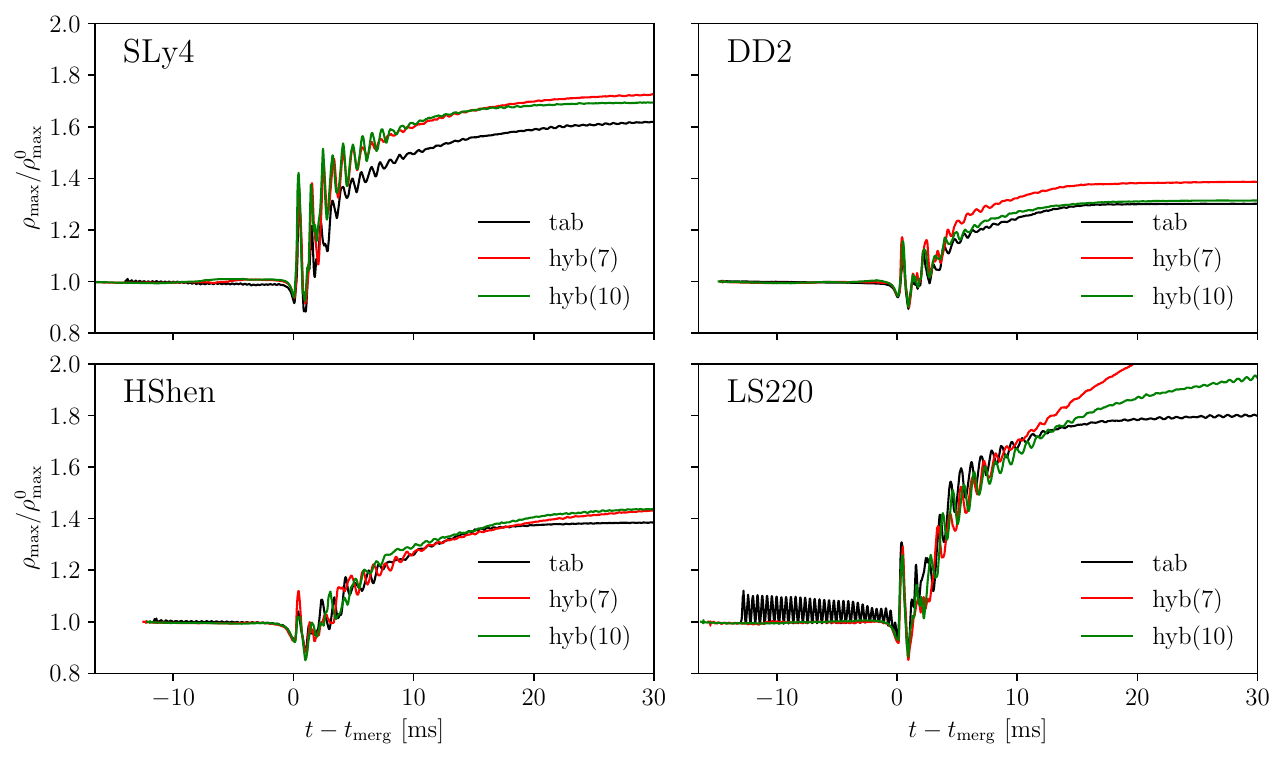}
\caption{Merger and early post-merger evolution of the maximum density (normalized by its initial value) for all four EOSs. Black solid lines indicate tabulated EOSs, red solid lines the $7-$pieces hybrid EOS, and green solid lines the $10-$pieces hybrid EOSs. Time is given with respect to the time of merger.}
\label{fig:rhomax}
\end{center}
\end{figure*}

Regarding the grid structure for the simulations, we employ a mesh refinement strategy using three grid centers, namely two of them at the center of each star and the third one at the center of the simulation domain. The use of multiple grid centers allows for higher resolution in regions of interest while keeping computational costs manageable~\cite{Schnetter:2003rb}. Each grid has five nested refinement levels, designed such that the innermost level contains $1.2$ times the radius of each star, with $45$ points along the radius. This setup ensures adequate resolution of the stellar interiors and the strong-field region. In this way, we obtain a mean resolution of $\Delta x = 240\,\rm m$ in the innermost refinement level, which is sufficient to capture the relevant physical processes during the merger~\cite{Bernuzzi:2011aq}. The outermost grid extends sufficiently far from the center (up to $1750$ km, on average) to properly capture GW emitted by the BNS systems, which are extracted at a finite radius using the Newman-Penrose formalism~\cite{Newman:1962qr}. 


\section{Results: early post-merger}
\label{sec:results_early}

\subsection{Matter evolution and GW emission}
\label{sub:waves}

\begin{figure}
\begin{center}
\includegraphics[width=1.\columnwidth]{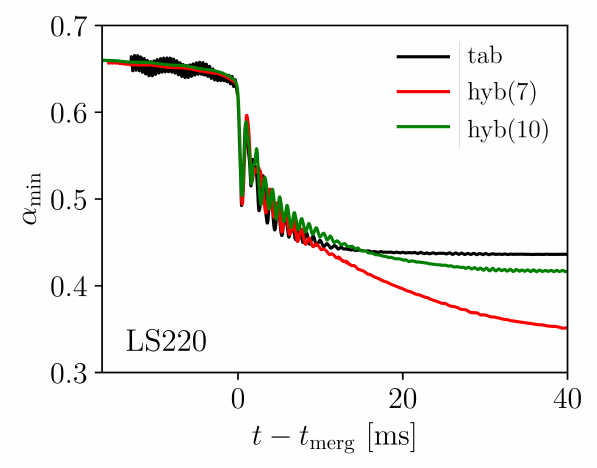}
\caption{Minimum of the lapse $\alpha_{\rm min}$ as a function of re-scaled time as in Fig~\ref{fig:rhomax} for the {\tt LS220} case. Lines and colors are the same as in Fig.~\ref{fig:rhomax}.}
\label{fig:minlapse}
\end{center}
\end{figure}

\begin{figure*}
\begin{center}
\includegraphics[width=2.\columnwidth]{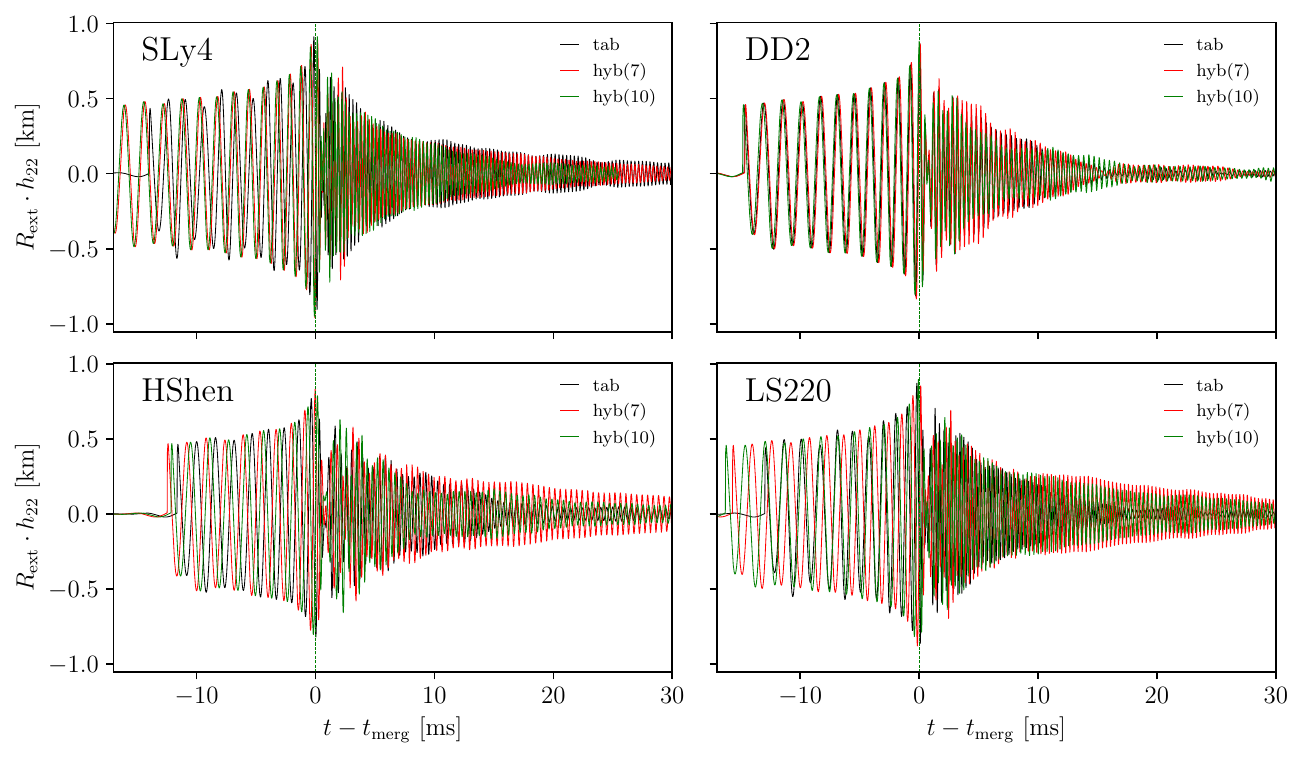}
\caption{$h_{22}$ component of the GW strain for all models. The colors for each curve follow the same criteria as in Fig.~\ref{fig:rhomax}.}
\label{fig:h22}
\end{center}
\end{figure*}

Fig.~\ref{fig:rhomax} shows the time evolution of the maximum density for all four EOSs. Time is expressed in $t-t_{\rm merger}$ and the figure only depicts the first $30\, \rm ms$ after merger. The evolution of the density offers a starting point to distinguish the difference in the dynamics between the hybrid and tabulated representations of the EOSs. We observe that the tabulated cases reach a lower value for the maximum density than the hybrid models. This indicates that the different treatment of thermal effects (and, hence, of thermal pressure), impact the way the remnant can sustain the gravitational pull~\cite{Sekiguchi2011,Bauswein2010}. Furthermore, the initial post-merger oscillations of the remnant  ($<10 \,\rm ms$) also differs in both approaches. For the {\tt DD2}, {\tt HShen} and {\tt LS220} EOSs, the oscillations are comparable in the two cases up to $\sim 10\,\rm ms$, while the {\tt SLy4} EOS presents differences already after $3\, \rm ms$. Comparing the models with $7-$pieces (red lines) and $10-$pieces (green lines), we observe that the {\tt SLy4} and {\tt HShen} cases exhibit similar behavior, reaching comparable central densities in the post-merger stage. In contrast, the other two EOSs yield different results. For the {\tt DD2} EOS, the tabulated data evolution overlaps with that of the $10-$pieces hybrid EOS, while in the {\tt LS220} case the central density increases until the remnant collapses to form a BH in both hybrid scenarios. Remarkably, the evolution of the {\tt LS220} model in the tabulated case does not lead to BH formation within our simulated time.

The evolution of the minimum value of the lapse function for model {\tt LS220} is plotted in Fig.~\ref{fig:minlapse}. The lapse decreases monotonically in all three cases. While for the tabulated EOS and the $10-$pieces hybrid EOS the slope of the drop almost plateaus, yielding a value of $\alpha_{\rm min}$ slightly above 0.4, the evolution of the $7$-piece hybrid case shows a rapid descent towards zero, signalling BH formation. An apparent horizon is indeed found in that case at $t-t_{\rm merger}\sim 66.2$ ms. We note that the $10-$pieces {\tt LS220} hybrid EOS will also likely result in the collapse of the post-merger remnant to a BH, yet on a longer timescale than we could afford to simulate. It is worth noting that, as the evolution of the the lapse suggests, the eventual collapse will not be in accordance with the model described by the  tabulated {\tt LS220} EOS. This is consistent  with previous studies~\cite{Radice2018,Perego:2017fho}.

Figure \ref{fig:h22} displays the GW signals from all simulations during the first $30$ ms post-merger. The reader is addressed to~\cite{Maione_2016, Maione_2017, De_Pietri_2016, DePietri2020} for details on the numerical extraction of the GW signal. The results show variations across the EOS sample and across the EOS representation of thermal effects. For {\tt SLy4} the signals of both the tabulated and $7-$pieces hybrid models overlap while the $10-$pieces model shows a greater damping of the postmerger amplitude. This behavior may be attributed to the differences in the high-density behavior captured by the piecewise polytropic fits~\cite{Read2009, Takami2015}. For the {\tt DD2} EOS the waveforms perfectly overlap during inspiral. However, after merger all three EOS representations exhibit a quasi-linear decay in the GW amplitude, at different rates. This continues until $15\ \rm ms$ post-merger, when the $7-$pieces hybrid and tabulated models reach a minimum amplitude, whereas the $10-$pieces model remains higher. This suggests that the detailed structure of the EOS influences the post-merger GW signal, consistent with findings in~\cite{Bernuzzi2015, Rezzolla2016}. The $7-$pieces hybrid {\tt HShen} EOS shows a longer linear decay than the other two, with the $10-$pieces model following closely the tabulated one. Lastly, the GW amplitude of the $7-$pieces and $10-$pieces hybrid {\tt LS220} is remarkably similar. Both mantain high amplitude even at $30\ \rm ms$ post-merger, while the tabulated model decays much more rapidly. This discrepancy may again arise from the inclusion of thermal effects in both representations~\cite{Kastaun2017, Endrizzi2018}.

\begin{figure*}
\begin{center}
\includegraphics[width=2.\columnwidth]{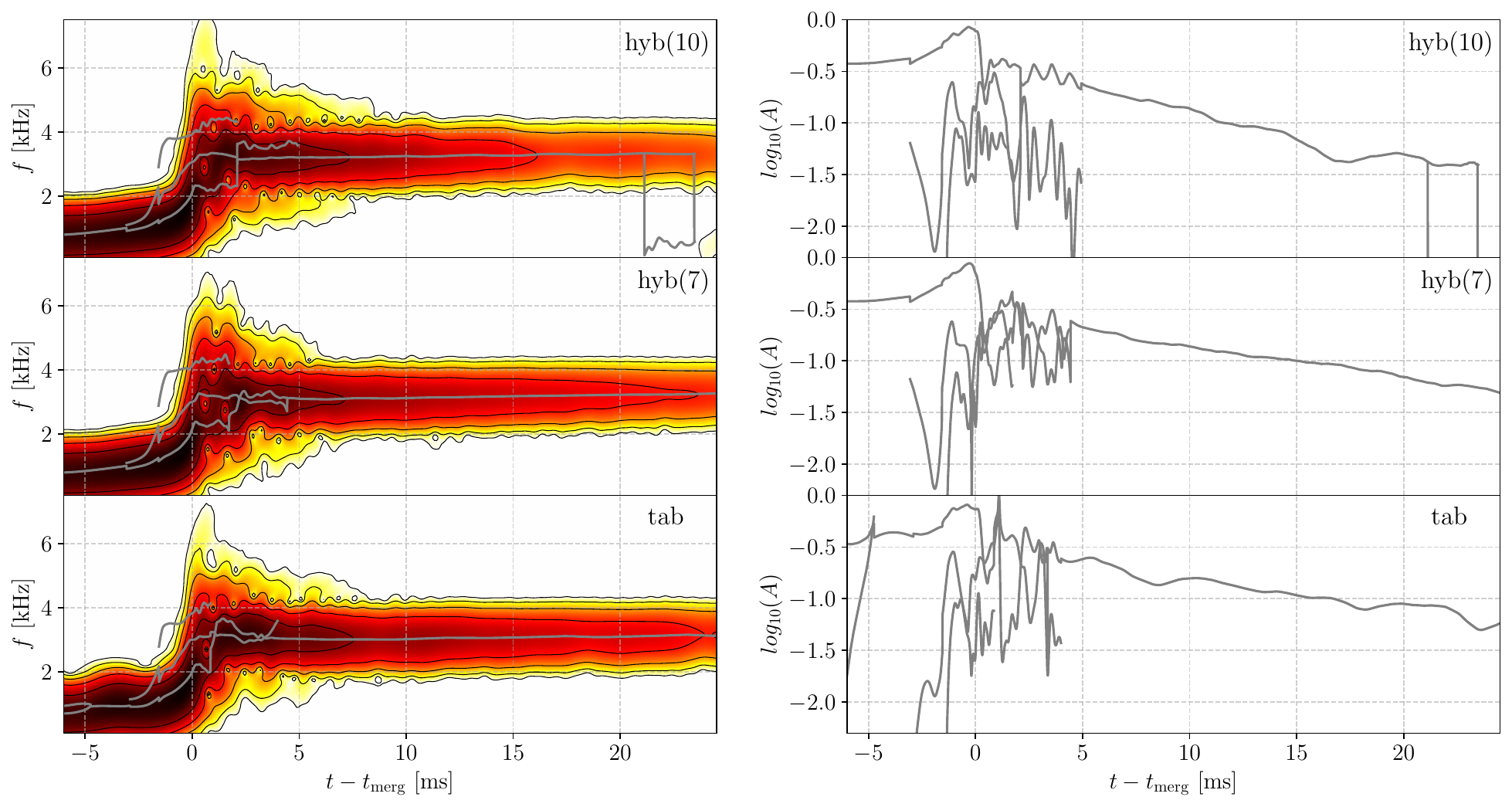}
\caption{Left panels: Time-frequency spectrograms of the $\ell=m=2$ component of the GW strain for the {\tt SLy4} model, as shown in Ref.~\cite{DePietri2020}. From top to bottom there are depicted the $10-$pieces hybrid case (top), the $7-$pieces hybrid case (middle) and the tabulated case (bottom). Thick gray lines, derived from the Prony analysis, indicate the frequency of the active modes primarily responsible for GW emission at various times. The spectral density's relative intensity is represented by color, with darker areas indicating higher intensity. Right panels: Amplitude of the main active modes (in arbitrary units).}
\label{fig:sly4spectro}
\end{center}
\end{figure*}

We observe notable differences analyzing the merger times, extracted from the GW strain, for the four different EOSs {\tt SLy4}, {\tt DD2}, {\tt HShen}, and {\tt LS220}. For the soft EOS {\tt SLy4} and {\tt LS220}, the merger times increase significantly from the tabulated to the piecewise models: {\tt SLy4} shows an increase from $13.39$ ms (tabulated) to $17.67$ ms ($7-$pieces) and $17.77$ ms ($10-$pieces), while {\tt LS220} increases from $12.98$ ms (tabulated) to $15.70$ ms ($7-$pieces) and $16.29$ ms ($10-$pieces). In contrast, the stiff EOSs {\tt DD2} and {\tt HShen} exhibit smaller variations: {\tt DD2} has merger times of $14.71\ \rm ms$ (tabulated), $14.66\ \rm ms$ ($7-$pieces), and $14.84\ \rm ms$ ($10-$pieces), and {\tt HShen} has $11.64\ \rm ms$ (tabulated), $12.47\ \rm ms$ ($7-$pieces), and $12.16\ \rm ms$ ($10-$pieces). The larger discrepancies in the soft EOSs can be attributed to their higher compactness and sensitivity to tidal deformations, which may not be as accurately captured by the hybrid representations~\cite{Damour2012, Hinderer2010}. The stiff EOSs, being less compact with larger tidal deformability, are better approximated by the hybrid models, resulting in smaller differences in merger times.

\subsection{Spectral analysis and quasi-universal relations}
\label{sub:universal}

The left panels of Fig.~\ref{fig:sly4spectro} show time-frequency plots of the $h_{22}$ component of the GW signal's spherical harmonic decomposition. We only present the {\tt SLy4} case to illustrate the analysis. Thick gray lines overlaid on the spectrogram represent the time evolution of the main active spectral modes of the remnant, determined using the Estimation of Signal Parameters via Rotational Invariance Techniques (\textsc{ESPRIT}) Prony’s method with a $3$ ms moving window, as discussed in~\cite{Maione_2017,DePietri2020}. Prony's method is a powerful technique for analyzing damped oscillatory signals by fitting a sum of exponential functions to the data~\cite{Takeda2017,Rezzolla2016,Breschi2019}. In the context of GW signals from BNS mergers, it allows for the extraction of quasi-normal mode frequencies and damping times of the post-merger remnant~\cite{Takeda2017,Breschi2019}. By applying this method, we can decompose the complex GW signal into its constituent modes, providing insights into the dynamics of the merger and the BNS remnant.

The peaks observed in the GW signal correspond to the dominant oscillation modes of the HMNS formed after merger~\cite{Rezzolla2016,Breschi2019}. These modes are sensitive to the EOS of dense nuclear matter, and their accurate identification may place constraints on the EOS~\cite{Breschi2019}. In particular, the fundamental $f$-mode and its overtones are prominent features in the post-merger GW spectrum, and their frequencies are correlated with NS properties such as radius and compactness~\cite{Takeda2017}. The corresponding mode amplitudes are displayed in the right panels of Fig.~\ref{fig:sly4spectro} in arbitrary units. The modes we are interested in are the so called $f_{2,i}$ mode which represent the dominant one in the first $\sim6-7$ ms after merger (\textit{early postmerger phase}), and the one in the remaining time $6-7$ ms $< t-t_{\rm merger}<30-40$ ms (\textit{intermadiate postmerger phase}), which is called $f_2$ in the literature~\cite{Rezzolla2016}. \\

\begin{figure*}
\begin{center}
\includegraphics[width=2.\columnwidth]{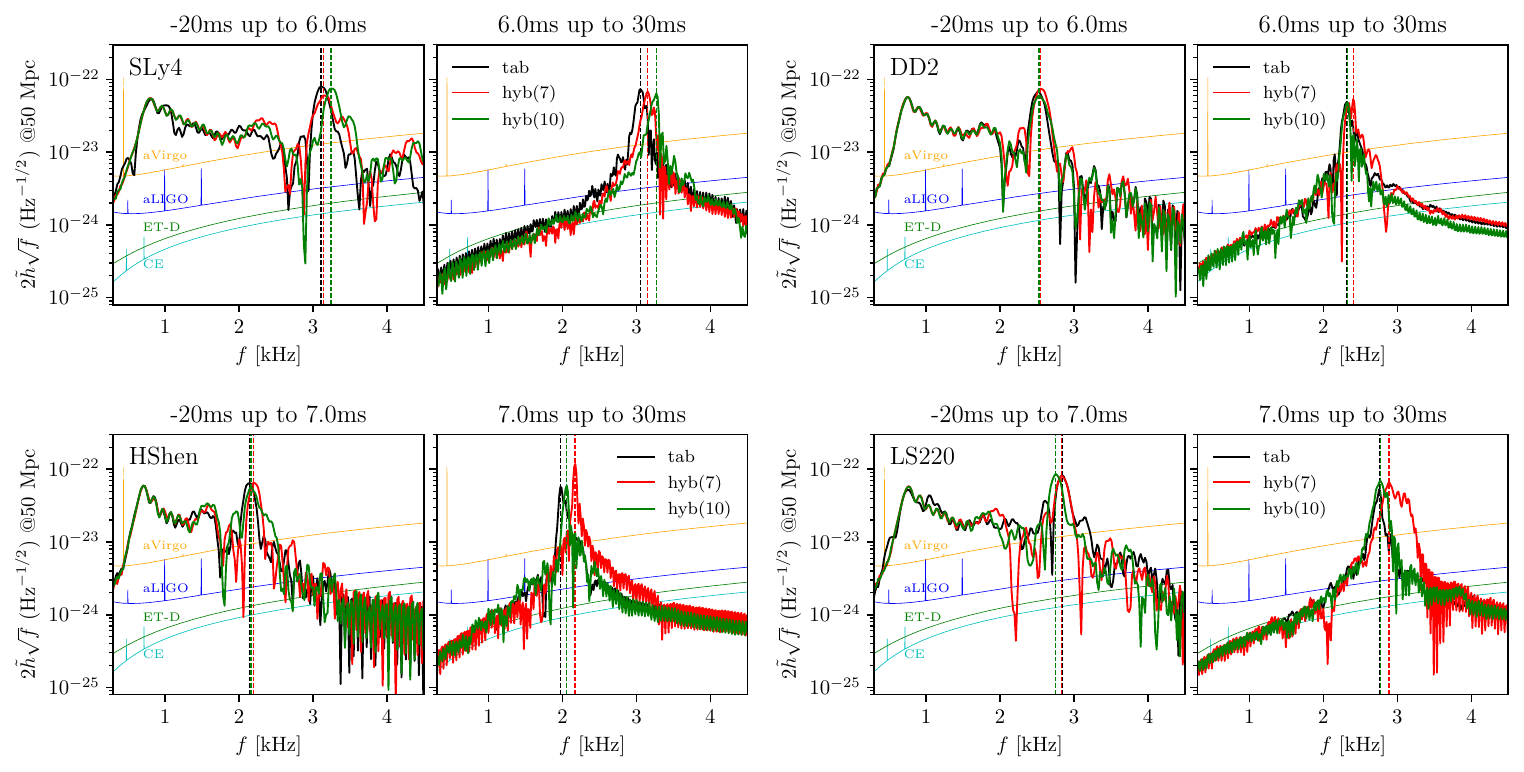}
\caption{GW spectra of BNS mergers at $50$ Mpc with optimal orientation for each EOS model. The black color represents the tabulated cases, while the $7-$ and $10-$pieces for the hybrid cases are depicted respectively in red and green. To emphasize the contribution of the prominent spectral components at different times, the spectra are given for two restricted time windows (marked at the top of each frame). The frequencies are expressed in kHz. Finally, the design sensitivities of Advanced LIGO~\cite{LIGOsens}, Advanced Virgo~\cite{VIRGO:2014yos}, Einstein Telescope~\cite{2011CQGra..28i4013H} and Cosmic Explorer~\cite{2017CQGra..34d4001A} are displayed.}
\label{fig:fftdiff}
\end{center}
\end{figure*}

\begin{center}
  \begin{table}[th]
    \caption{Main GW frequencies for the $f_{\rm max}$ mode (left), the $f_{2,i}$ mode (middle), and the $f_2$ mode (right), given in kHz. The notation {\tt hyb(number)} indicates the number of pieces used in the hybrid EOS version. The $f_2$ mode for the {\tt LS220} hybrid cases is not reported since this system later collapses to a BH and the frequency is increasing.
      \label{table:freqs}}
        \setlength{\tabcolsep}{8pt}
        \begin{tabular}{ccccl}
            \multicolumn{1}{l}{} & \multicolumn{1}{l}{} & \multicolumn{1}{l}{} & \multicolumn{1}{l}{} &  \\ \cline{1-5}
            EOS & $t_{\rm merg}$ & $f_{\rm max}$ & $f_{2,i}$ & \multicolumn{1}{c}{$f_2$} \\ \cline{1-5}
            {\tt SLy4}$_{\rm tab}$ & 13.98 & 1.85 & 3.16 & 3.16 \\
            {\tt DD2}$_{\rm tab}$ & 14.71 & 1.57 & 2.53 & 2.33 \\
            {\tt HShen}$_{\rm tab}$ & 11.64 & 1.44 & 2.16 & 1.99 \\
            {\tt LS220}$_{\rm tab}$ & 12.98 & 1.78 & 2.85 & 2.74 \\ \cline{1-5}
            {\tt SLy4}$_{\rm hyb(7)}$ & 17.67 & 1.91 & 3.15 & 3.15 \\
            {\tt DD2}$_{\rm hyb(7)}$ & 14.66 & 1.57 & 2.53 & 2.35 \\
            {\tt HShen}$_{\rm hyb(7)}$ & 12.47 & 1.48 & 2.20 & 2.18 \\
            {\tt LS220}$_{\rm hyb(7)}$ & 15.70 & 1.71 & 2.84 & - \\ \cline{1-5}
            {\tt SLy4}$_{\rm hyb(10)}$ & 17.77 & 1.92 & 3.24 & 3.27 \\
            {\tt DD2}$_{\rm hyb(10)}$ & 14.84 & 1.59 & 2.53 & 2.32 \\
            {\tt HShen}$_{\rm hyb(10)}$ & 12.15 & 1.40 & 2.16 & 2.06 \\
            {\tt LS220}$_{\rm hyb(10)}$ & 16.29 & 1.73 & 2.75 & - \\ \cline{1-5}
            \multicolumn{1}{l}{} & \multicolumn{1}{l}{} & \multicolumn{1}{l}{} & \multicolumn{1}{l}{} & 
        \end{tabular}
  \end{table}
\end{center}

Fig.~\ref{fig:fftdiff} shows the GW spectra of an optimally oriented BNS merger at $50$ Mpc for all EOS models. Black lines correspond to the tabulated case while the hybrid cases with $7-$ and $10-$pieces are represented in red and green, respectively. To highlight the contribution of key spectral components at different times, the spectra are presented for two specific time windows, marked at the top of each frame. Thus, the maximum peak in each time window represents the $f_{2,i}$ mode for the first window and the $f_2$ mode for the last window. Design sensitivities for Advanced LIGO~\cite{LIGOsens}, Advanced Virgo~\cite{VIRGO:2014yos}, Einstein Telescope~\cite{2011CQGra..28i4013H}, and Cosmic Explorer~\cite{2017CQGra..34d4001A} are included to show the possible signal detectability. A summary of the values for these frequencies is reported in Table~\ref{table:freqs}. This table also includes the value of $f_{\rm max}$, corresponding to the instantaneous GW frequency at the time of merger. For detailed definitions of these frequencies we refer the reader to~\cite{Topolski:2023ojc}. \\

Our results show that adding more segments into the hybrid representation of the EOS does not necessarily yield to mode frequencies closer to those obtained with the  tabulated version of the EOS. This happens for {\tt HShen}, where both the $f_{2,i}$ and $f_2$ modes with the $10-$pieces approximation are closer to the tabulated frequency than the $7-$pieces hybrid version. Similarly, for {\tt DD2}, no significant differences are observed across the cases. Conversely, {\tt SLy4} shows a larger frequency difference in the $10-$pieces hybrid case compared to the $7-$pieces case, as does the $f_{2,i}$ mode frequency for the {\tt LS220} case. All EOSs but {\tt DD2}  exhibit frequency differences as large as about $50$ Hz for the $f_{2,i}$ mode and around $200$ Hz for the $f_2$ mode, which are greater than the differences observed for $f_{\rm max}$. 

We use the extracted mode frequencies to build \textit{quasi-universal relations}, empirical relations between macroscopic properties of NSs that exhibit a weak dependence on the EOS. These relations allow to infer key NS properties from observational data without precise knowledge of the EOS~\cite{Bauswein:2011tp,Rezzolla2016,Rezzolla:2017aly}. In Ref.~\cite{Rezzolla2016}, using the tidal polarizability parameter $\kappa^{\rm T}_2$,
\begin{equation}
    \kappa_2^T\equiv \frac{1}{8} \bar{k}_2 \left( \frac{\bar{R}}{\bar{M}} \right)^5 = \frac{3}{16} \Lambda \,,
\end{equation}
where $\bar{k}_2$ is the dimensionless tidal Love number, $\bar{R}$ is the average radius of the NS in the binary system, $\bar{M}$ is the average mass, and $\Lambda= \frac{2}{3}k_2 ({\bar{M}}/{\bar{R}})^{-5}$ denotes the dimensionless tidal deformability, a quasi-universal relation which connects the instantaneous frequency at merger with NS properties was reported,
\begin{equation}
\label{eq:fmaxRezzolla}
    \log_{10} \left( \frac{f_{\rm max}}{\text{Hz}} \right) \approx a_0 + a_1 \left( \kappa^{\rm T}_2 \right)^{1/5} - \log_{10} \left( \frac{2 \bar{M}}{M_\odot} \right)\,.
\end{equation}
Here, $a_0$ and $a_1$ are empirical fit coefficients calibrated from numerical simulations of equal-mass BNS mergers. This relation has been recently extended by~\cite{Topolski:2023ojc} for the unequal-mass case. It shows that a larger mass results in a lower peak frequency, reflecting the faster dynamics of more massive BNS mergers.

\begin{figure}
\begin{center}
\includegraphics[width=1.\columnwidth]{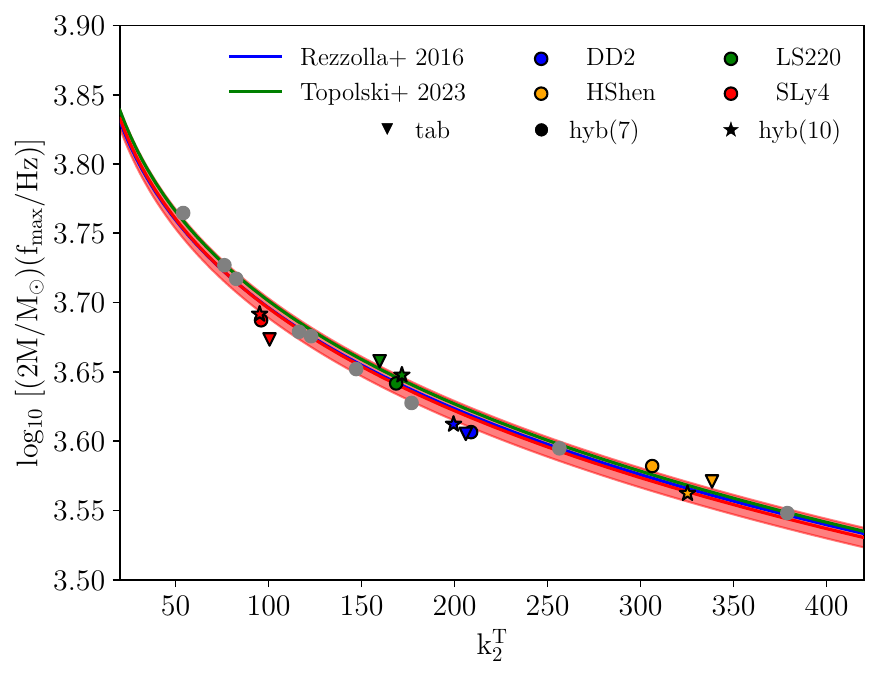}
\caption{Mass-weighted frequencies at the merger time $f_{\rm max}$ shown as a function of the tidal polarizability parameter $\kappa^{\rm T}_2$. Filled triangles of different colours refer to the tabulated models, colored circles (stars) to the $7-$pieces ($10-$pieces) hybrid ones, and gray circles represent the hybrid cases simulated in~\cite{Rivieccio_2024}. The red solid line shows the fit as given by Eq.~\eqref{eq:fmaxRezzolla} (with our $a_0$ and $a_1$ coefficients) while the blue and green solid lines represent the fit obtained respectively by~\cite{Rezzolla2016} and~\cite{Topolski:2023ojc}. The red region indicates $1\sigma$ (standard deviation) from our fit.}
\label{fig:fmax}
\end{center}
\end{figure}

Fig.~\ref{fig:fmax} shows the results of our empirical fit for the $f_{\rm max}$ frequency as a function of $\kappa^{\rm T}_2$ alongside those from~\cite{Rezzolla2016,Topolski:2023ojc}. The values for our fit coefficients are $a_0 = 4.194 \pm 0.020$ and $a_1 = -0.198 \pm 0.007$, with a standard deviation $\sigma\approx 0.0071$. The fit is calculated using only hybrid cases with a $7-$pieces approximation to ensure consistency with the data from~\cite{Rivieccio_2024}, generated using the same initial masses and separation and evolved with the same code. These additional data, shown with gray circles in the figure, increase the number of points for the fit. Notably, during the early post-merger the frequencies from both tabulated and hybrid cases closely follow our fit and those reported in previous works, deviating by less than $2\sigma$, as illustrated by the $1\sigma$ standard deviation displayed in the plot. We note that deviations from the fit for tabulated and hybrid models may be also affected by differences in the tidal polarizability parameter between the two models at the level of initial data. This parameter is highly sensitive to the chosen gravitational mass and circumferential radius of a single NS, resulting in significant variations from small differences in the EOS initial data.

\begin{figure}
\begin{center}
\includegraphics[width=1.\columnwidth]{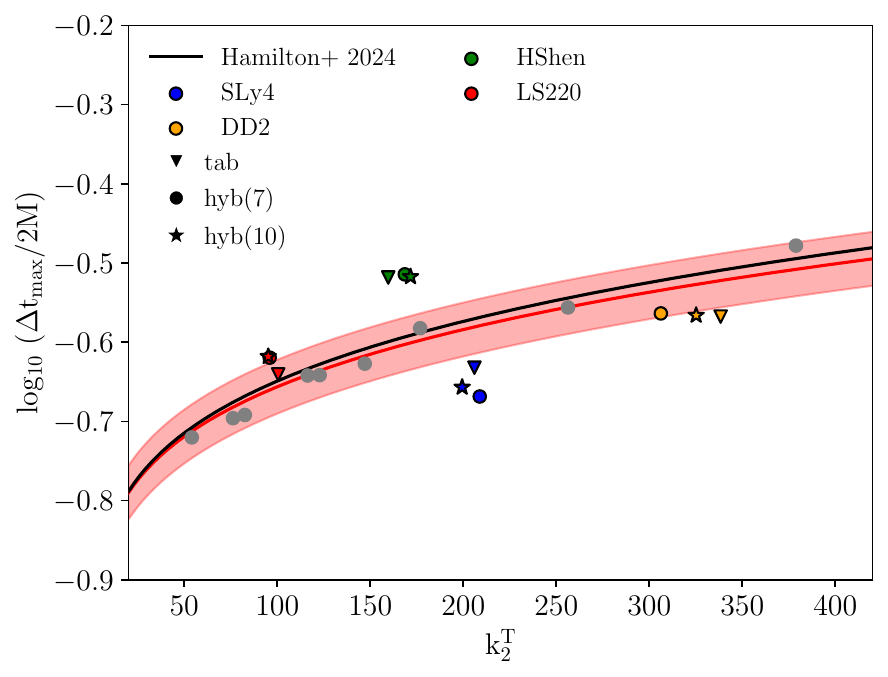}
\caption{Time difference between the time of merger and the time when peak instantaneous frequency is attained as a function of the tidal polarizability parameter $\kappa^{\rm T}_2$. Our fit is shown in red with its $1\sigma$ standard deviation shaded, while the fit from~\cite{Hamilton:2024ziw} is shown in black. Data symbols have the same meaning as in Fig~\ref{fig:fmax}. }
\label{fig:tmax}
\end{center}
\end{figure}
 
A new quasi-universal relation linking the interval between the time of merger and the time when the maximum instantaneous frequency is attained, $\Delta t_{\rm max} = t_{\rm merger} - t_{\rm max}$, has recently been reported by~\cite{Hamilton:2024ziw},
\begin{equation}
    \Delta t_{\rm max} = a_0 + a_1 \left( \kappa^{\rm T}_2 \right)^{1/5} \rm kHz\ .
\end{equation}
Using this empirical fit with our data we obtain the following coefficients, $a_0 = -1.14 \pm 0.11$, and $a_1 = 0.193 \pm 0.040$, and a standard deviation of $\sigma\approx 0.034$. Our results are depicted in Fig.~\ref{fig:tmax}, where a comparison with the fit from~\cite{Hamilton:2024ziw} is also shown. As in~\cite{Hamilton:2024ziw} some outliers to the fit are observed. Overall, we find broad agreement between our fit and that from~\cite{Hamilton:2024ziw} and not significant differences between hybrid and tabulated models, the {\tt LS220} and {\tt DD2} EOSs being the closest ones to the fit.\\

\begin{figure}
\begin{center}
\includegraphics[width=1.\columnwidth]{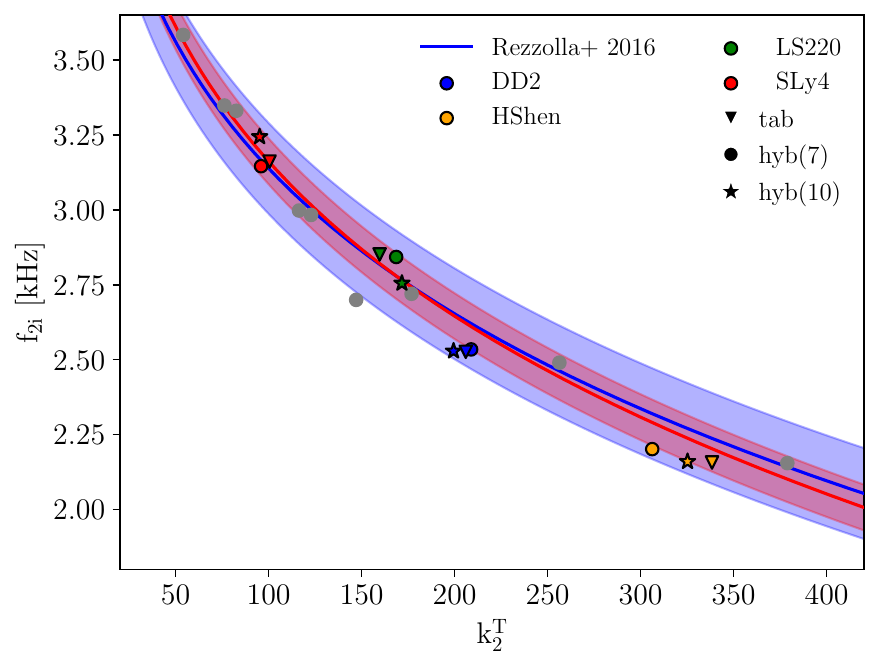}
\caption{Frequencies at the early postmerger phase, $f_{2,i}$, as a function of $\kappa^{\rm T}_2$. The data and fits are as in Fig.~\ref{fig:fmax}. The colored regions represent the $1\sigma$ standard deviation for each fit.}
\label{fig:f2i}
\end{center}
\end{figure}

\begin{figure}
\begin{center}
\includegraphics[width=1.\columnwidth]{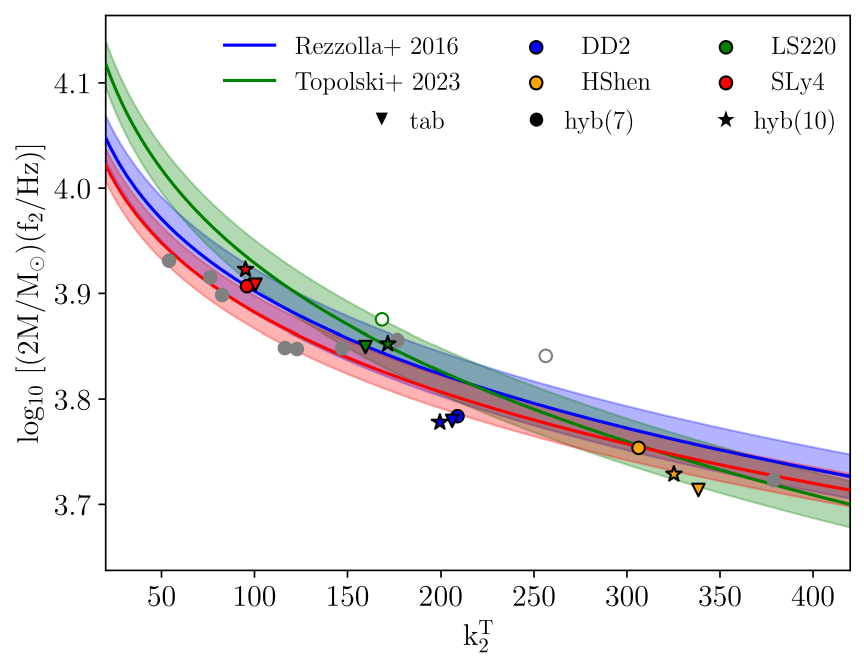}
\caption{Mass-weighted frequencies at the intermediate post-merger phase, $f_2$, as a function of $\kappa^{\rm T}_2$. The data and fits are as in Fig.~\ref{fig:fmax}. The colored regions represent the $1\sigma$ standard deviation for each fit.}
\label{fig:f2}
\end{center}
\end{figure}

For the $f_{2,i}$ frequency, which is the dominant mode in the first $6-7$ ms after merger, we use the fitting formula from~\cite{Rezzolla2016},
\begin{equation}
    f_{2,i} = a_0 + a_1 \left( \kappa^{\rm T}_2 \right)^{1/5} \rm kHz \ ,
\end{equation}
with coefficients $a_0 = 6.636 \pm 0.148$, $a_1 = -1.383 \pm 0.055$, and a standard deviation $\sigma\approx 0.077$. The results are shown in Fig~\ref{fig:f2i}, where we compare our findings to those in~\cite{Rezzolla2016}. Here, differences between models begin to emerge. Those become more apparent for the next dominant mode $f_2$, as illustrated in Fig.~\ref{fig:f2}, as the system evolves further away from the merger. At this stage, discrepancies between the two EOS representations become more evident. For {\tt HShen} and {\tt DD2} (stiff EOS), modes $f_{2,i}$ and $f_2$ exhibit lower frequencies than expected based on the quasi-universal relations. In contrast, {\tt SLy4} and {\tt LS220} (soft EOS) display higher frequencies than predicted. Our results also indicate that frequency shifts between models based on hybrid and tabulated EOS increase further away from the merger. These discrepancies could impact analyses of real GW events that rely on quasi-universal relations to extract information about the EOS structure~\cite{Miravet2024,VillaOrtega2024}. 

\section{Results: late post-merger}
\label{sec:latepostmerger}

We turn now to discuss the late post-merger phase of the remnants. A particular focus of this analysis is to assess how the incorporation of finite-temperature effects through tabulated EOSs affects the possible excitation of inertial modes in the HMNS. The late-time appearance of those modes was first reported by~\cite{DePietri:2018tpx,DePietri2020}, albeit for simulations restricted to hybrid EOSs. Here, we will compare results from both tabulated and hybrid EOS, using only $7-$pieces approximations for the latter, with simulations extending well beyond $100$ ms post-merger.

Analysing the evolution of the maximum of the matter density we observe that the stiff EOS models ({\tt DD2} and {\tt HShen}) yield less compact remnants than the soft EOS ones ({\tt SLy4} and {\tt LS220}), as expected from the previous analysis. Moreover, the comparison between tabulated and hybrid models reveals that the latter produce more compact remnants. Specifically, the compactness increases in the hybrid models by $10.5\%$ for {\tt DD2}, $10.2\%$ for {\tt HShen}, and $13.0\%$ for {\tt SLy4}. Furthermore, for the {\tt LS220} EOS, the hybrid model collapses to a BH at $t_{\rm BH}=66.2 \,\rm ms$ after merger, a phenomenon not observed in the tabulated case. This suggests that the hybrid approach has a greater impact on soft EOSs than on stiff ones.

\begin{figure*}
\begin{center}
\includegraphics[width=2.\columnwidth]{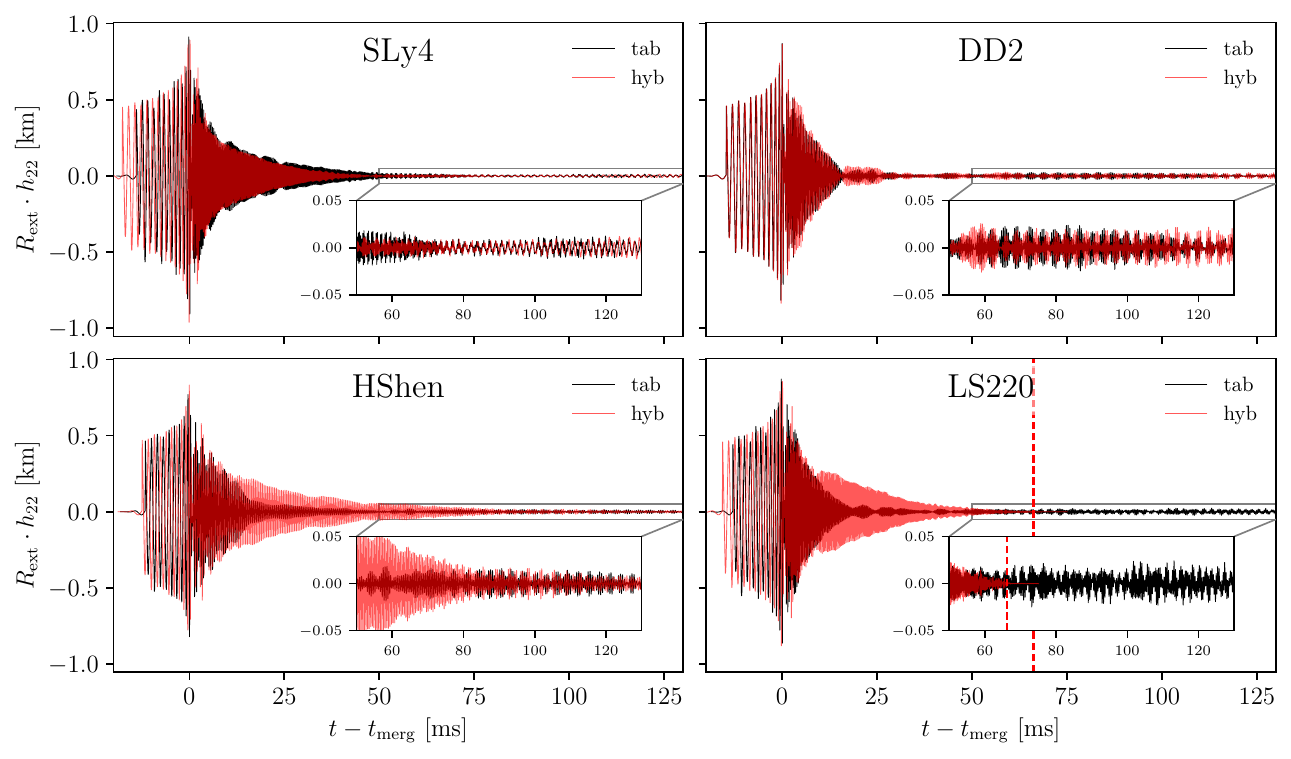}
\caption{Scaled $h_{22}$ component of the GW strain as a function of time (relative to the time of merger) for all models. Black lines indicate tabulated EOSs while red lines correspond to the hybrid counterparts (with 7 pieces in the polytropic part). The vertical red dashed line for the {\tt LS220}$_{\rm hyb}$ case marks the time of BH formation. Insets zoom in on late post-merger waveforms.}
\label{fig:h22_long}
\end{center}
\end{figure*}

\subsection{GW signal and spectral analysis}
\label{sub:lateGW}

A tabulated-vs-hybrid comparison of the GW signals for all models  is shown in Fig.~\ref{fig:h22_long}. This figure depicts the $h_{22}$ strains. As expected, for all models the inspiral part of the signal is hardly affected by the approach of thermal effects in the EOS. The differences become visible only in the post-merger evolution. In this part, the {\tt SLy4} EOS shows a rapid decline in amplitude over $\sim 50$ ms. The hybrid model does not exhibit strong oscillations during this decline, whereas the tabulated model shows amplitude modulations that persist throughout the simulation. The signal amplitude decays in a quasi-linear way during the first $20$ ms post-merger for the {\tt DD2} EOS. In this EOS both the tabulated and the hybrid case exhibit a modulation in the signal. As we will discuss next, this indicates the emergence of inertial modes. Similar behavior is present for the {\tt HShen} EOS, with the hybrid case exhibiting a slower, prolonged decay than the tabulated counterpart. Both signals, however, display a similar behavior toward the end of the simulation.

The models evolved with the {\tt LS220} EOS show the most significant differences between the two thermal approaches: the hybrid model collapses to a BH at $t_{\rm BH} = 66.2$ ms after merger, while the remnant of the tabulated counterpart  is stable against gravitational collapse for the duration of the simulation. This dynamics is reflected on the waveforms depicted in Fig.~\ref{fig:h22_long} where the tabulated EOS exhibits a stronger and faster decay post-merger, whereas the hybrid case shows a rise of the signal $10$ ms after merger. Moreover, in the tabulated case, modulations of the GW signal (indicating re-excitation) appear only after $20$ ms and persist throughout the signal, albeit at a significantly smaller amplitude than in the early post-merger part.

\begin{figure*}
\begin{center}
\includegraphics[width=2.\columnwidth]{PLOTS/emitted_percent.pdf}
\caption{Time evolution of GW emission quatities for the four EOSs. Each row corresponds to a different EOS, and columns display respectively: the energy (left), the angular momentum (center), and the linear momentum (right) emitted in GWs. All quantities are normalized by the late-time values obtained from the tabulated EOS simulations (shown in black). The corresponding hybrid EOS models are shown in red. Time is given relative to the merger time.}
\label{fig:emittedGW}
\end{center}
\end{figure*}

To quantify the differences in GW emission between tabulated and hybrid EOSs, we compute the total energy $E_{\rm GW}$, angular momentum $J_{\rm GW}$, and linear momentum $P_{\rm GW}$ radiated as a function of time. These quantities are obtained by integrating the time derivatives of the multipolar components of the Weyl scalar $\psi_4$, extracted at a finite radius, over time. The radiated energy rate is computed from the strain, evaluating the standard energy flux expression~\cite{Ruiz:2007yx},
\begin{equation}
\frac{dE_{\rm GW}}{dt} = \frac{1}{16\pi} \sum_{\ell,m} \left| \dot{h}_{\ell m}(t) \right|^2\ ,
\end{equation}
where $h_{\ell m}$ is obtained via two time integrations of $\psi_4^{\ell m}$. 

Similarly, the angular momentum flux along the $z$-axis is calculated incorporating up to $\ell_{\text{max}} = 4$ modes and reconstructing the time derivative of the angular momentum as~\cite{Ruiz:2007yx},
\begin{equation}
\frac{dJ_{\rm GW}}{dt} = \frac{1}{16\pi} \sum_{\ell,m} m \, \operatorname{Im} \left[ \dot{h}_{\ell m}(t) \, \bar{h}_{\ell m}(t) \right]\ .
\end{equation}

The linear momentum flux is computed employing the full set of expressions detailed in~\cite{Ruiz:2007yx}, combining different $\ell$-mode interactions to obtain the recoil force components $dP^i_{\rm GW}/dt$. The total radiated quantities are then obtained by time integration:
\begin{equation*}
E_{\rm GW}(t) = \int_0^t \frac{dE_{\rm GW}}{dt'} dt' \ , \qquad
J_{\rm GW}(t) = \int_0^t \frac{dJ_{\rm GW}}{dt'} dt' \ ,
\end{equation*}
\begin{equation*}
P_{\rm GW}(t) = \sqrt{ \sum_{i = x,y,z} \left( \int_0^t \frac{dP^i_{\rm GW}}{dt'} dt' \right)^2 }\ .
\end{equation*}

\begin{figure*}
\begin{center}
\includegraphics[width=2.\columnwidth]{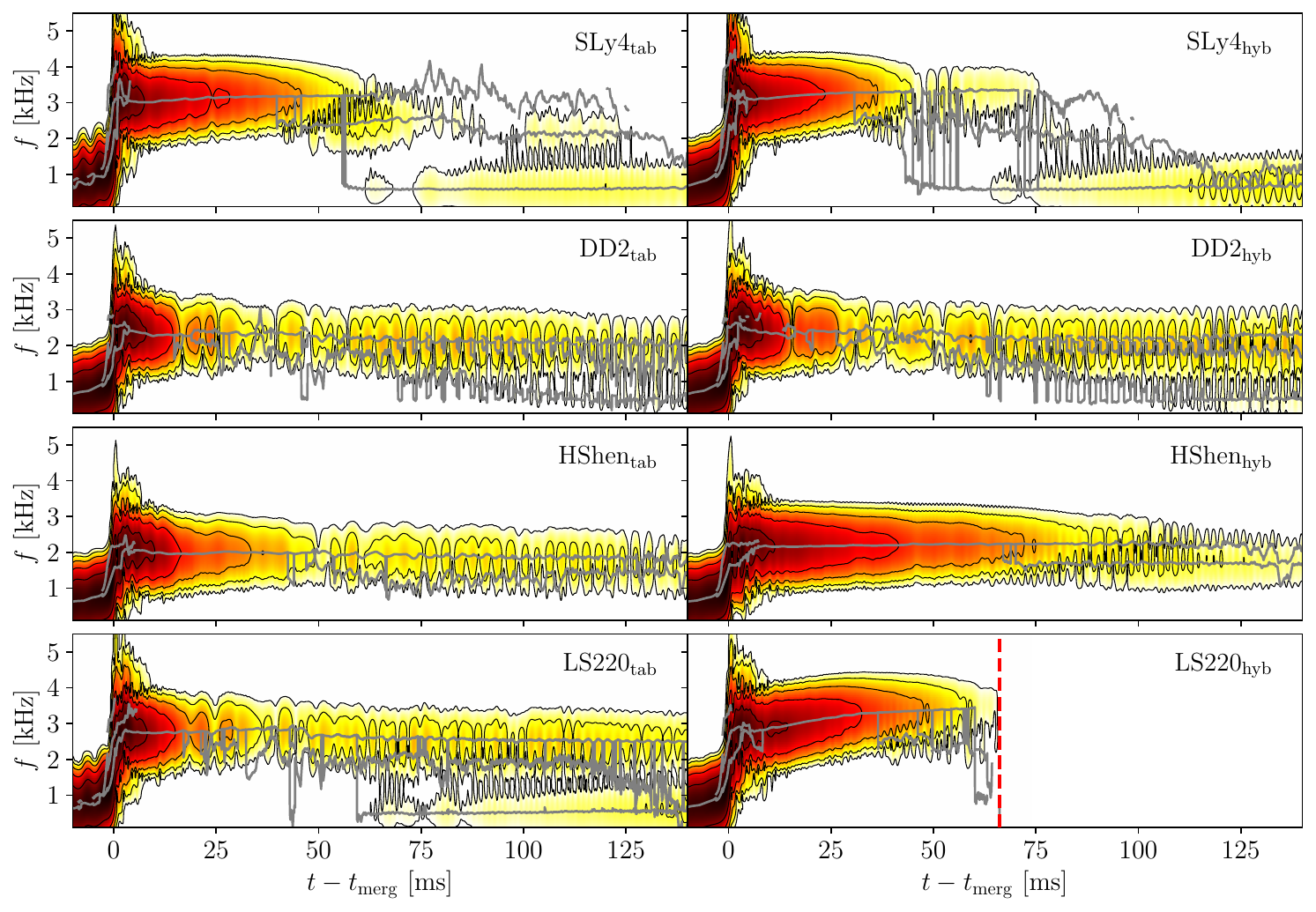}
\caption{Time-frequency spectrograms of the $\ell=m=2$ component of the GW strain for all models. Each row depicts a distinct EOS, with the left panel showing tabulated cases and the right panel displaying hybrid cases. Thick grey lines indicate the frequency of the active modes responsible for GW emission at various times, as determined by Prony's analysis. The relative intensity of the spectral density is represented by color, with darker areas indicating higher intensity. The dashed red vertical line indicates BH formation for the hybrid {\tt LS220} model.}
\label{fig:prony_long}
\end{center}
\end{figure*}

Fig.~\ref{fig:emittedGW} depicts these quantities, normalized to the late-time values from the tabulated EOS simulation for direct comparison with hybrid models. For most EOSs, hybrid models exhibit higher energy and linear momentum emission. The {\tt HShen} case shows the largest difference, with the hybrid model emitting $45.56\%$ more energy $E_{\rm GW}$ and $30.60\%$ more angular momentum $J_{\rm GW}$ compared to the tabulated model. In contrast, hybrid approaches emit less linear momentum $P_{\rm GW}$ than tabulated ones, except for the {\tt SLy4} EOS. In this case, the tabulated approach emits slightly less energy and a similar amount of angular momentum but significantly less linear momentum. 

We turn now to discuss the GW spectral analysis, following the approach laid out in~\cite{DePietri2020}. Fig.~\ref{fig:prony_long} displays the spectrograms using Prony's method with a $3\,\rm ms$ time window applied to the entire GW signal. Grey lines in this figure indicate the frequency of active GW modes. A common behavior across all EOSs is that, at a certain time, the grey lines exhibit a jump from the $f_2$ mode frequency, dominant in the early post-merger, to a lower frequency mode. As we discuss in detail below, this indicates the emergence of inertial modes. 

For the {\tt DD2} EOS the spectrogram shows minimal differences between the tabulated and hybrid cases, with the most obvious difference being the collapse of the remnant to a BH in the latter. In the {\tt SLy4} case, differences are observed not only in the intensity of the spectrogram but also in the timing of the appearance of the second mode, at $t-t_{\rm merg}\sim 40\,\rm ms$ for the tabulated EOS and $t-t_{\rm merg}\sim 30\,\rm ms$ in the hybrid case. The dominant $f_2$ frequency, represented by a nearly straight grey line for the first $40\,\rm ms$, is lower in the tabulated case compared to the hybrid one. At late post-merger, both models show a low-frequency oscillation below $1\,\rm kHz$. This is also visible for the {\tt LS220} and {\tt DD2} EOSs, although is less pronounced in the latter. The cause of this low-frequency oscillation remains uncertain; it may result from numerical errors induced by the behavior of the gauge~\cite{Ciolfi_2019} or it might be a physical non-axisymmetric mode excited due to the lack of system symmetries of our simulation setup. 

The major differences in the spectrogram are observed for the {\tt HShen} EOS. In this case, the tabulated model exhibits a lower average frequency than the hybrid case and the GW signal has weaker intensity (and amplitude). Modulations in the spectrogram appear starting at $50\,\rm ms$ post-merger in the tabulated case, while in the hybrid case, they emerge only at $90\,\rm ms$. This timing difference in the appearance of the inertial mode is also evident, around $40\,\rm ms$ in the tabulated model compared to $65\,\rm ms$ in the hybrid case.

\begin{figure*}[h!]
\begin{center}
\includegraphics[width=2.\columnwidth]{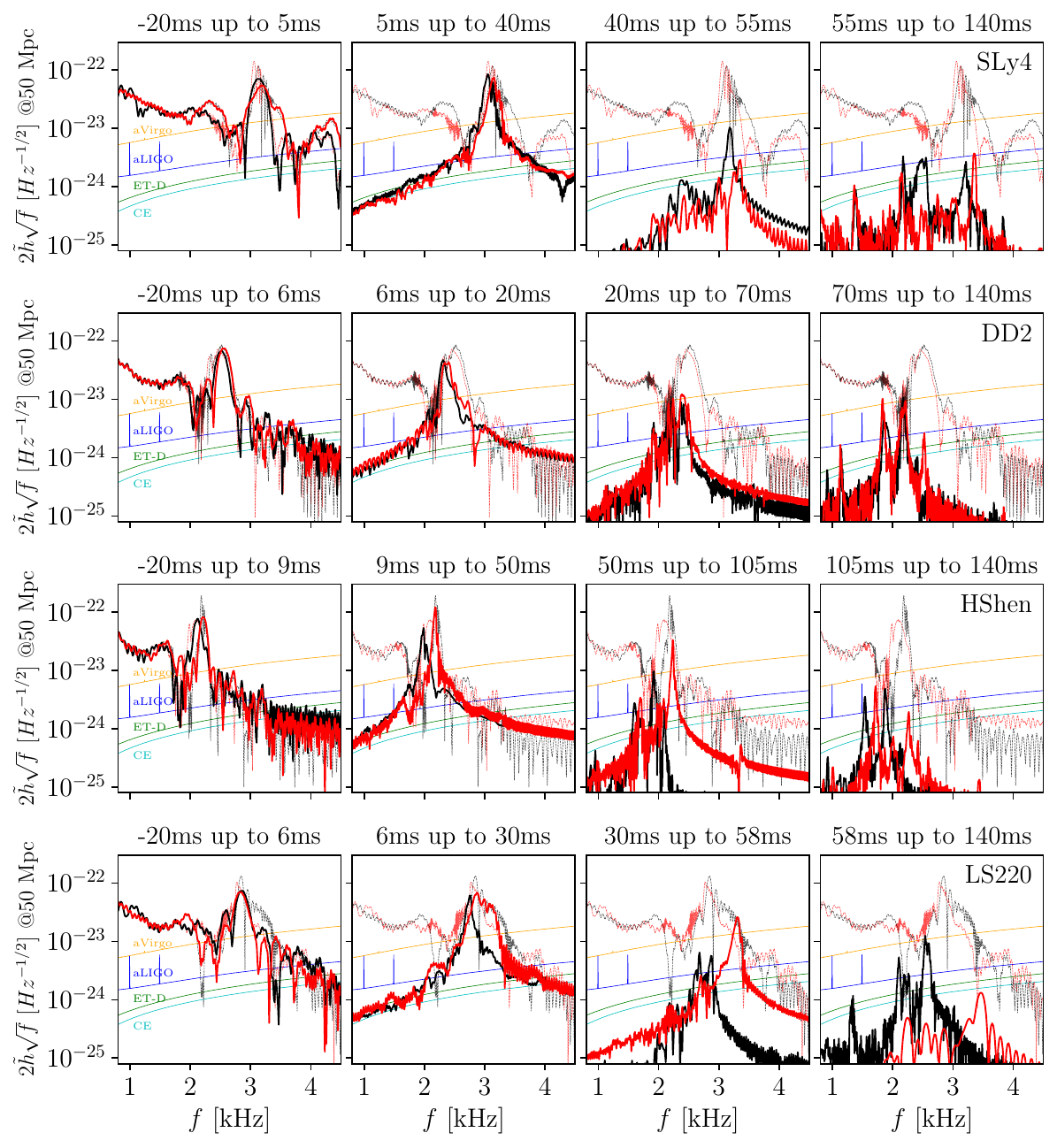}
\caption{GW spectra of BNS mergers at $50$ Mpc with optimal orientation. Each row corresponds to a different EOS and each column indicates a different time window (marked at the top of each frame). Black lines represent the tabulated EOSs while the hybrid ones are depicted in red. To emphasize the contribution of the spectral components at different times, the spectra are given for the complete GW signal (thin-dashed lines) and for four restricted time windows (thick-solid lines). The design sensitivities of Advanced LIGO~\cite{LIGOsens}, Advanced Virgo~\cite{VIRGO:2014yos}, Einstein Telescope~\cite{2011CQGra..28i4013H} and Cosmic Explorer~\cite{2017CQGra..34d4001A} are also displayed.}
\label{fig:fft_long}
\end{center}
\end{figure*}

Fig.~\ref{fig:fft_long} shows the spectra of the modes identified in the spectrograms in frequency space. We split the data into four spectra with restricted time windows, allowing us to emphasize each frequency mode's contribution to the GW signal. The figure also includes the design sensitivities of Advanced LIGO~\cite{LIGOsens}, Advanced Virgo~\cite{VIRGO:2014yos}, Einstein Telescope~\cite{2011CQGra..28i4013H}, and Cosmic Explorer~\cite{2017CQGra..34d4001A} to provide a perspective on observational possibilities. Dashed thin lines represent the spectra from the entire time window of the simulation, while thick solid lines are the spectra for each selected time window for the tabulated case (black) and the hybrid model (red). 

\begin{figure*}[t]
\begin{center}
\includegraphics[width=2.\columnwidth]{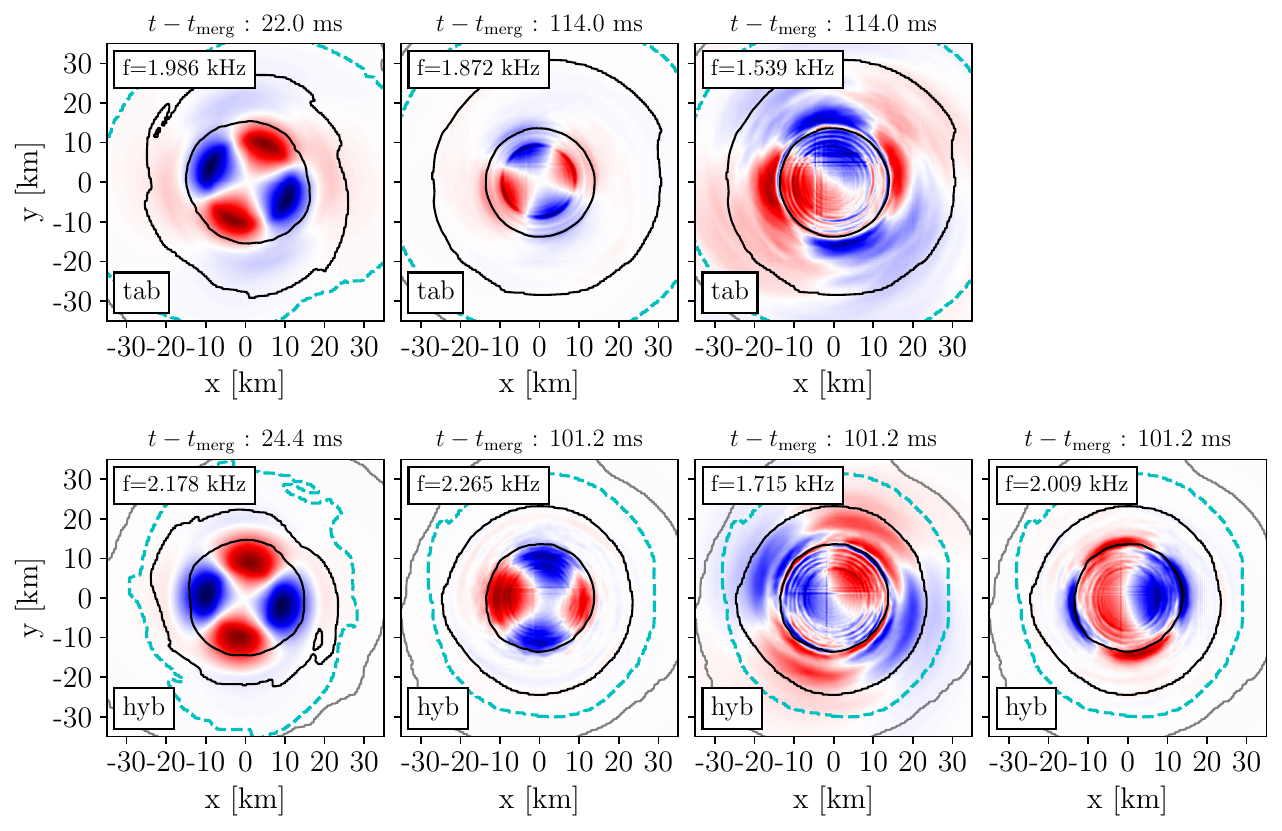}
\caption{Density eigenfunctions in the equatorial planes of the {\tt HShen} model. The top row corresponds to the tabulated case, while the bottom row represents the hybrid case. Each panel displays a snapshot at different times after the merger (noted at the top) to highlight when the relevant frequency (indicated in the top left corner) is most apparent. The black and gray lines denote equal-density contour lines at $\sim(10^{12},10^{13},10^{14})\,\rm g/cm^3$ and the cyan dashed line marks the core (bulk) of the remnant.}
\label{fig:eigen}
\end{center}
\end{figure*}

The frequency of the main modes is seen to shift between tabulated and hybrid models as the HMNS evolves. This shift suggests that thermal contributions during the early post-merger stage can influence the remnant's evolution in the late post-merger phase. In the {\tt SLy4} case, the $f_2$ mode shifts to higher frequencies, reaching $f_2^{\rm late}=3.21$ kHz (tabulated) and $f_2^{\rm late}=3.35$ kHz (hybrid)\footnote{In Sec.~\ref{sub:eigenfunction} we describe on how we identify the $f_2$ mode frequency peak, which persists until the end of the simulation.}. Additionally, an inertial mode emerges at a lower frequency, namely $f_{\rm inertial}=2.14$ kHz (hybrid) and  $f_{\rm inertial}=2.17$ kHz (tabulated). The latter case also shows a wide frequency peak at $2.5$ kHz, which is absent in the hybrid case, suggesting an additional mode that the hybrid approximation does not capture. 

The {\tt DD2} case demonstrates a similar behavior in both approaches: the tabulated approach yields $f_2^{\rm late}=2.17$ kHz and $f_{\rm inertial}=1.86$ kHz, while the hybrid case shows $f_2^{\rm late}=2.20$ kHz and $f_{\rm inertial}=1.83$ kHz. Notably, a third frequency peak at $2.52$ kHz appears only in the hybrid case. The largest frequency shifts are attained in the {\tt HShen} EOS. Both the late $f_2$ mode and the inertial mode are higher in the hybrid case ($f_2^{\rm late}=2.26$ kHz and $f_{\rm inertial}=1.71$ kHz) compared to the tabulated one ($f_2^{\rm late}=1.87$ kHz and $f_{\rm inertial}=1.54$ kHz). Similar to the {\tt DD2} case, the hybrid model features an additional frequency peak at $2.0$ kHz, positioned between the two main modes. 
Lastly, the late post-merger evolution of the {\tt LS220} EOS only shows frequency peaks in the tabulated case, as the hybrid model collapses to a BH ($f_2^{\rm late}=2.54$ kHz and $f_{\rm inertial}=2.10$ kHz, along with an additional peak at $1.34$ kHz).

In summary, the largest frequency shift occurs in the {\tt HShen} case, with the $f_2^{\rm late}$ mode shifted by $390\,\rm Hz$ between the two approaches. The observed frequency shifts and additional peaks can be attributed to the differences in how thermal and compositional effects are treated in the tabulated and hybrid EOS models. In the hybrid model, where thermal contributions are approximated, the post-merger remnant's thermal evolution is less accurately captured, leading to higher $f_2$ frequencies in the late phase. The extra peaks in the tabulated models, especially for {\tt DD2}, indicate additional modes inadequately captured by the hybrid approximation. While some of these modes correspond to peaks in other components ($\ell = 2,m=1$ and $\ell = 2,m=0$), as seen in the {\tt DD2} EOS, others, such as the extra peak for {\tt HShen} EOS, lack corresponding peaks in other components. These modes likely arise from more complex interactions between thermal pressure, composition, and the star's structure, which the tabulated EOS better resolves. Addressing this would require further investigation.

\subsection{Mode analysis: eigenfunctions}
\label{sub:eigenfunction}

We now focus on the spatial structure of the observed modes by analyzing their \textit{eigenfunctions}. This provides insight into the density and temperature distributions of the remnant as it oscillates. The extracted eigenfunctions correspond to the dominant oscillation modes, namely the fundamental quadrupole $m=2 \ f$-mode and the inertial mode. The former typically exhibits a simple structure, particularly in the equatorial plane, where it shows no nodal lines, highlighting its fundamental nature. On the other hand, the inertial mode has a different pattern in the equatorial plane, presenting different radial nodes. 

\begin{figure*}
\begin{center}
\includegraphics[width=2.1\columnwidth]{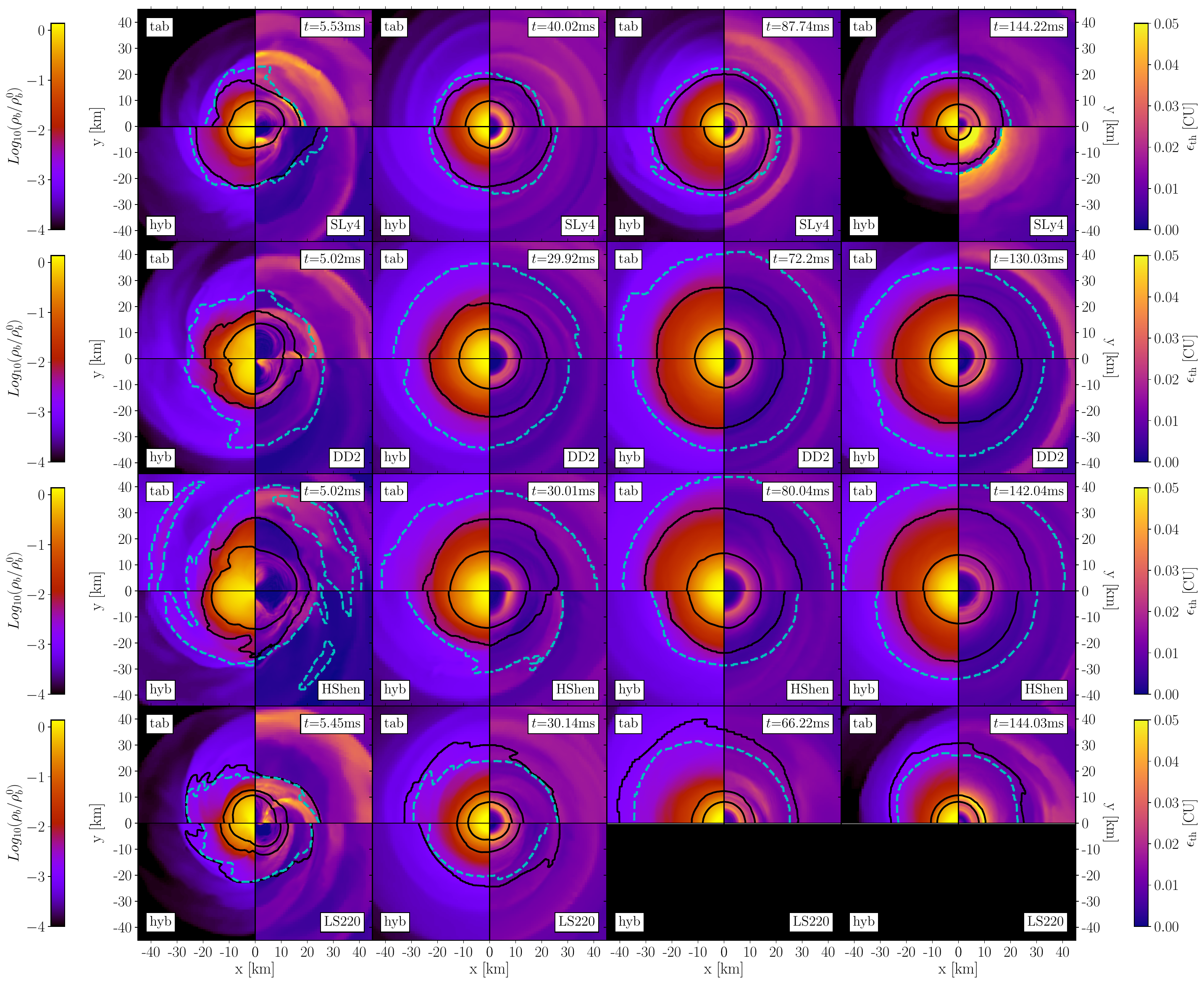}
\caption{Snapshots for different times (columns) on the equatorial plane of normalized rest-mass density in logarithmic scale (left-half part of each panel), and the thermal energy in code units (right-half part of each panel). Each row represents a different EOSs, indicated in the bottom right corner, while the top-half part of each panel displays the tabulated models and the corresponded hybrid model in the bottom-half part. The black solid lines are equal-density contour lines at $\sim(10^{12},10^{13},10^{14})$ g/cm$^3$. Finally, the cyan dotted line shows the surface of the bulk of the remnant.}
\label{fig:snapshots}
\end{center}
\end{figure*}

Fig.~\ref{fig:eigen} depicts the eigenfunctions for the {\tt HShen} EOS. We focus on this case as the results are likewise across all EOSs. The figure shows density eigenfunctions in the equatorial ($z=0$) plane for the tabulated model (top) and the hybrid case (bottom) at various post-merger times. Each panel is normalized to the maximum density eigenfunction, with positive values in red and negative values in blue. The leftmost panel illustrates the $f_2$ mode, characterized by a quadrupolar shape without radial nodal lines, which aids in tracking this mode during the late post-merger phase depicted in the second column. Isocontours of matter density are indicated by black and grey lines, while the cyan line marks the bulk of the HMNS, calculated as the isocontour where the maximum matter density decreases by three orders of magnitude. (Further details will be discussed in Section~\ref{sub:temperature}.) The intensity of the $f_2$ mode in the late post-merger phase is in agreement with the results from  Fig.~\ref{fig:h22_long} and Fig.~\ref{fig:prony_long},  where the GW signal amplitude for the {\tt HShen} EOS differs between the tabulated and hybrid models, the former exhibiting lower values. This trend is mirrored in the eigenfunction, with the tabulated case showing a weaker intensity than the hybrid case.

The eigenfunction structure of the inertial mode is visible in the last two columns of Fig.~\ref{fig:eigen}. It displays a quadrupolar shape with a nodal radial line where the eigenfunction changes sign. The excited low-frequency modes in the late post-merger phase exhibit features compatible with inertial modes. Their frequencies are appropriately ranged, with the dominant mode consistently having a frequency slightly lower than the star's maximum rotational frequency (see discussion in Section~\ref{sub:rotation}). Despite our limited EOS sample size, it appears that the dominant mode frequency in this phase correlates with the star's rotational frequency, as expected for inertial modes. 

\subsection{Thermal energy and temperature}
\label{sub:temperature}

\begin{figure*}[t]
\begin{center}
\includegraphics[width=2\columnwidth]{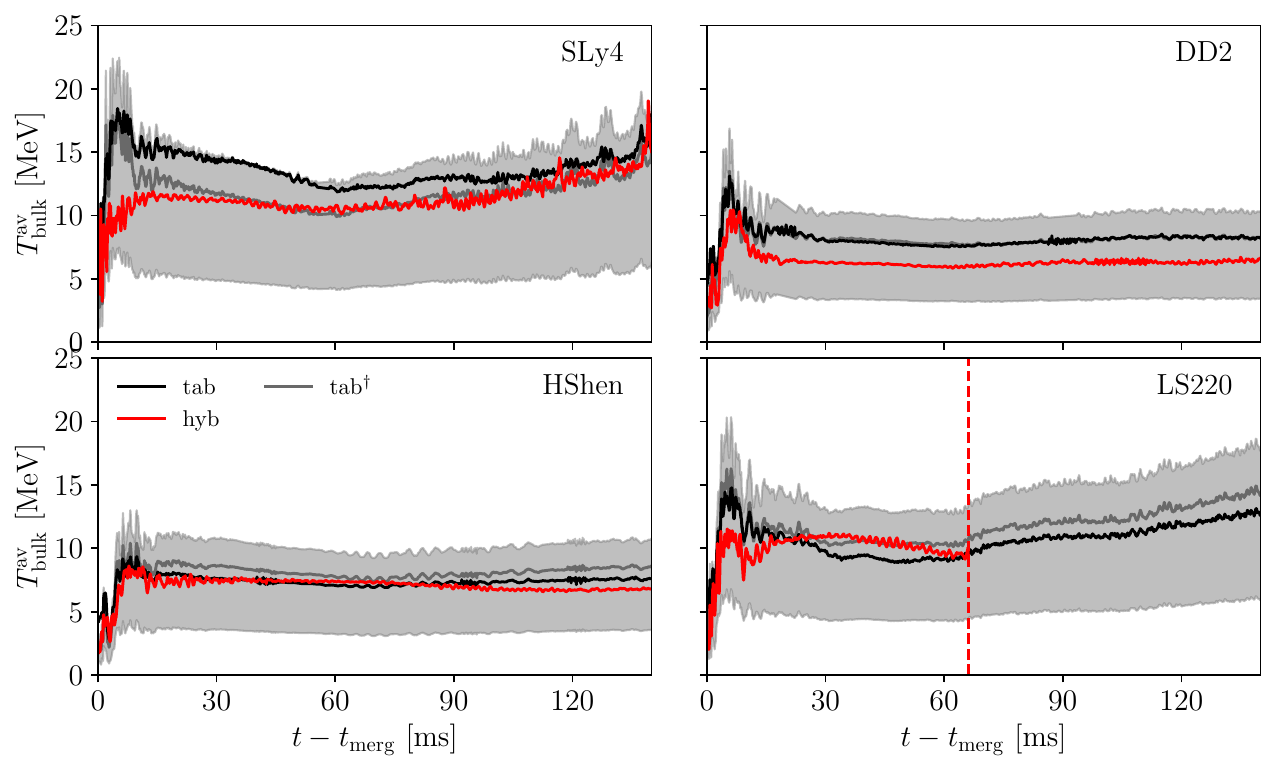}
\caption{Time evolution of the average temperature inside the bulk of the remnant for tabulated EOSs (black lines) and hybrid ones (red lines). Grey lines indicate the temperature for tabulated EOSs but evaluated as for hybrid EOSs using Eq.~\eqref{eq:hybridtemp} with $\Gamma_{\rm th}=1.8$, while the gray region indicates the region obtained by evaluating the latter with $4/3<\Gamma_{\rm th}<2$. The vertical red dashed line indicates the time of BH formation for the hybrid {\tt LS220} EOS.}
\label{fig:bulktemp}
\end{center}
\end{figure*}

Before exploring the origin of the inertial mode and its connection to the remnant's stability, we examine the matter density and thermal energy distribution in the equatorial plane. Fig.~\ref{fig:snapshots} displays these quantities for each EOS at various post-merger times (noted in the top right corner), illustrating the remnant's characteristics during the early, intermediate, and late post-merger phases.  The rest-mass density is shown on a logarithmic scale normalized to its maximum value at $t=0$ (left half of each panel), while the thermal energy is represented in code units (right half of each panel). We report thermal energy instead of temperature because the latter cannot be uniquely defined for the hybrid EOS, whereas it is directly obtained from the original EOS table. Further discussion can be found in Appendix~\ref{app:Temperature}. Each row corresponds to a specific EOS while the columns represent different post-merger times. Moreover, each panel is divided into two horizontal sections: the upper section displays the tabulated case and the lower section corresponds to the hybrid counterpart. The black solid lines indicate equal-density contours at $\sim(10^{12},10^{13},10^{14})\,\rm g/cm^3$, and the cyan dotted line represents the region containing most of the remnant, defined as the isocontour where the maximum rest-mass density decreases by three orders of magnitude. This setup allows us to observe the remnant's extension along the equatorial plane. 

Figure~\ref{fig:snapshots} highlights both common and differing features among the EOS models. A common key aspect is that the average thermal energy distribution in the tabulated cases is generally higher than in the hybrid ones. This may be attributed to the different temperature characterizations: tabulated EOS better captures temperature dynamics as temperature is a variable within the tables, while the hybrid case approximates thermal energy, potentially leading to inaccurate temperature estimates. This pattern holds across various models, although for the {\tt SLy4} EOS, thermal energy in the hybrid case increases at later simulation times, surpassing that of the tabulated case. Furthermore, this difference is also present between soft and stiff EOS. For the soft EOS {\tt SLy4} and {\tt LS220}, the thermal energy in the surrounding regions reaches higher values compared to the stiff EOS {\tt DD2} and {\tt HShen}. 

Several factors may explain this behavior. First, soft EOSs are generally more compressible, allowing for greater density changes under pressure. This increased compressibility may lead to more significant shock heating in the outer layers of the remnant, as material is compressed more efficiently during the merger~\cite{Bauswein:2011tp}. In contrast, stiff EOSs, which are less compressible, may limit the amount of heating in the outer layers, resulting in comparatively lower thermal energy in the surrounding material. Second, soft EOSs tend to produce remnants with lower central densities and higher temperatures in the outer layers due to stronger thermal pressure contributions at a given density. This can lead to an enhanced release of thermal energy, while stiff EOSs, dominated by cold nuclear pressure, may radiate less thermal energy in the outer regions~\cite{Radice2018b}. Finally, soft EOS may induce more vigorous turbulence and shock dynamics in the outer regions due, again, to their compressibility, which can further enhance the dissipation of kinetic energy into thermal energy. This results in higher radiated thermal energy compared to the more stable dynamics of stiff EOS remnants~\cite{Bauswein:2011tp}. Another potential contributing factor worth mentioning is the efficiency of neutrino cooling, enhanced in stiff EOSs~\cite{Radice2018b}. This effect, not included in our modelling, will be investigated elsewhere.

Figure~\ref{fig:snapshots} also shows the emergence of a hot ring at the center of the remnant for all models, noticeable $30\,\rm ms$ after the merger, when most energy has radiated away and the system transitions from a bar-shaped morphology to a more spherical shape~\cite{Kastaun:2016yaf}. This spherical phase is visible in the second snapshot for all models except the {\tt HShen} hybrid case, which requires more time to dissipate energy. The presence of this hot ring is closely related with the development of convective instabilities in the remnant, as we discuss in Section~\ref{sub:stability} below.

A final notable feature is the size of the remnant where hybrid EOSs produce smaller remnants than tabulated EOSs. The most significant difference occurs in the {\tt HShen} case, with an average disparity of nearly $10$ km. This behavior may stem from the better thermal energy description in the tabulated EOS, which provides greater thermal pressure to counteract gravitational forces, resulting in a larger remnant. 

The differences in thermal characterization are evident in the average temperature within the star's bulk, as shown in Fig.~\ref{fig:bulktemp}. The panels display the average temperature for each EOS in the remnant. An important distinction arises in how temperature is defined. For tabulated EOSs, temperature is interpolated directly from the tables, resulting in the black lines in Fig.~\ref{fig:bulktemp}. In contrast, for hybrid EOSs, temperature is approximated using the equation~\cite{DePietri2020},
\begin{equation}
    \label{eq:hybridtemp}
    T_{\rm hyb} = (\Gamma_{\rm th} - 1)\epsilon_{\rm th}\, .
\end{equation}
Here, $\Gamma_{\rm th}$ corresponds to the one in the polytropic part used in Eq.~\eqref{eq:pthermal}. The hybrid EOS temperature is represented by the red line in the figure. In Appendix~\ref{app:Temperature}, we provide a critical assessment of this approximation and suggest that it underestimates the real temperature of degenerate fluid elements. Given this analysis, we also calculate the tabulated case temperature using Eq.~\eqref{eq:hybridtemp} with $\Gamma_{\rm th}=1.8$, shown in gray. In addition, the gray region in the plot represents the latter definition using the theoretical expected region for $\Gamma_{\rm th}$, where $4/3<\Gamma_{\rm th}<2$. By comparing these three different ways to define the temperature, we observe that the average temperature of the remnant computed using the last two definitions exhibits a similar behavior only for the {\tt DD2}  where the grey and black lines almost coincide. On the other hand, for the {\tt HShen} and {\tt LS220} EOSs the temperature calculated using Eq.~\eqref{eq:hybridtemp} overestimates the actual temperature computed using the tables while for the {\tt SLy4} EOS we see a lower estimation for the temperature. Furthermore, the value of the $\Gamma_{\rm th}$ that recovers the original tabulated temperature is $1.72$ for the {\tt HShen} and {\tt LS220} EOSs, and between $1.90$ and $1.98$ for the {\tt SLy4} case.  

At the time of merger, the temperature in the remnant reaches values exceeding $80\,\rm{MeV}$ across all EOSs. However, at later times, the maximum temperature measured in the bulk of the remnant typically settles in the range of $30$–$40\,\rm{MeV}$ for the tabulated models. When comparing this with the temperature computed using the hybrid EOS or the tabulated EOS but adopting the hybrid definition in Eq.~\eqref{eq:hybridtemp}, we consistently find an underestimation of about $10$–$15\,\rm{MeV}$ in the final stages of the evolution consistent with the analysis in Appendix~\ref{app:Temperature}. It is important to note that the mentioned maximum temperature refers to the peak value reached within the bulk of the remnant, which may correspond to just a few grid zones. Further investigation is needed to assess the spatial extent and robustness of these high-temperature regions.\\
Our analysis underscores the importance of accurately defining temperature, particularly when comparing tabulated and hybrid EOSs. The differences in thermal energy characterization, driven by the inherent limitations of the hybrid approach, may explain key disparities in the remnant's evolution, such as the variation in size and thermal dynamics. These differences also highlight the role of the EOS stiffness in determining remnant properties. 

\subsection{Convective instabilities in the remnant}
\label{sub:stability}

\subsubsection{Brunt-Väisälä frequency}
\label{sub:stability}

We now discuss the issue of the stability of the HMNS against convection. This stability is mainly influenced by buoyancy forces shaped by the star's internal stratification. To assess the local convective stability of the post-merger remnant we analyze the \textit{Brunt–Väisälä frequency}, $\mathcal{N}^2$, which quantifies the restoring force acting on a fluid element displaced from equilibrium. Following~\cite{Shibata:2025nfj} we assume that the remnant has an axisymmetric shape and adopt cylindrical coordinates $(r, \phi, z)$, where $r$ is the cylindrical radius from the rotation axis. Assuming the HMNS is approximately in hydrostatic equilibrium and axisymmetric after $ \sim 20{-}30$ ms post-merger, the Brunt–Väisälä frequency for radial displacements for incompressible perturbations in the equatorial plane simplifies to
\begin{equation}
\mathcal{N}^2 = \gamma_{rr}^{-1/2} \, G_r \, B_r\ ,
\label{eq:BVfreq}
\end{equation}
where $\gamma_{rr}$ is the cylindrical component of the spatial metric and the quantity $G_r$ is the \textit{gravitational acceleration} defined as,
\begin{equation}
    G_r = (e + P)^{-1} \, \partial_{r} P\ ,
\label{eq:shibataG}
\end{equation}
where $e$ is the total energy density, and $P$ is the pressure. The quantity $B_r$ is the radial component of the \textit{Ledoux discriminant}, defined as
\begin{equation}
B_{r} = \frac{1}{e + P} \, \partial_{r} e - \frac{1}{\Gamma_1 P} \, \partial_{r} P\ ,
\label{eq:Avarpi}
\end{equation}
where $\Gamma_1 $ is the adiabatic index for adiabatic perturbations, incorporating both cold and thermal contributions to the EOS.
This expression emerges from a local linear perturbation analysis under the short-wavelength approximation, in which perturbations vary over scales much shorter than the size of the star. The analysis neglects metric perturbations, assuming the stability is governed by local fluid behavior. The focus on the radial direction arises from the assumption that the largest gradients in pressure and entropy, key drivers of convective stability, occur in the $r$-direction in axisymmetric rotating bodies. This simplification is justified in the equatorial plane, where deviations from equilibrium are most relevant for instability development. 

A more accurate definition of the gravitational acceleration, $G_r$, that does not assume hydrostatic equilibrium is reported in~\cite{Torres-Forne:2017xhv,Tseneklidou:2025fid}, 
\begin{equation}
    G_r = - \partial_r\, \ln{\alpha}\ ,
\label{eq:dimiG}
\end{equation}
where $\alpha$ is the lapse function. This expression reduces to the hydrostatic equilibrium case described by Eq.~(\ref{eq:shibataG}) in the absence of rotation. 

\begin{figure}
\centering
\includegraphics[width=1\columnwidth]{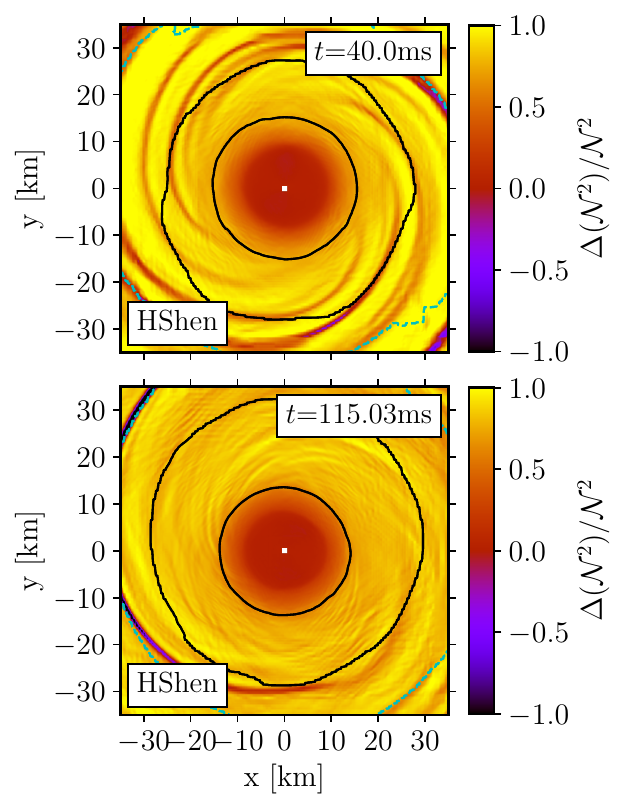}
\caption{Relative error $\Delta(\mathcal{N}^2)/\mathcal{N}^2$ in the equatorial plane of the remnant for the {\tt HShen} tabulated EOS, computed at $t-t_{\rm merg} = 40\,\rm ms$ (top panel) and $115.03\,\rm ms$ (bottom panel). The error quantifies the deviation in the Brunt–Väisälä frequency when using the hydrostatic approximation in Eq.~\eqref{eq:shibataG} instead of the definition from Eq.~\eqref{eq:dimiG}.}
\label{fig:BVerror}
\end{figure}

To quantify the implications of using either approach on the convective stability analysis, we compute the relative error in the Brunt–Väisälä frequency for all our EOS models, defined as 
\begin{equation*}
\frac{\Delta(\mathcal{N}^2)}{\mathcal{N}^2} = \frac{\mathcal{N}^2 - \mathcal{N}_P^2}{\mathcal{N}^2}\ ,
\end{equation*}
where $\mathcal{N}^2$ is computed using Eq.~\eqref{eq:dimiG} and $\mathcal{N}_P^2$ with Eq.~\eqref{eq:shibataG}. Fig.~\ref{fig:BVerror} displays this quantity in the equatorial plane for the {\tt HShen} tabulated EOS at two representative times, $t = 40\,\rm ms$ and $t = 115.03\,\rm ms$ after merger. The results indicate that the relative error remains moderate ( $\lesssim 10\%$ ) within the innermost $\sim 5$--$7\,\rm km$ from the center, suggesting that the hydrostatic equilibrium assumption remains a reasonably good approximation in the core. This could be related to the nearly rigid rotation in this inner region of the remnant (see Sec.~\ref{sub:rotation}). However, the error rapidly increases in the intermediate and outer layers of the remnant, reaching $\sim 50\%$ beyond $\sim10\,\rm km$ and values close to unity after $\sim20\,\rm km$ in the outer spiral arms. 

This analysis demonstrates that Eq.~\eqref{eq:shibataG} systematically underestimates the gravitational acceleration compared to the more general expression in Eq.~\eqref{eq:dimiG}, especially in the outer layers of the remnant where significant thermal and rotational gradients drive convective dynamics. In particular, while the two definitions differ in magnitude, they do not alter the sign of the Brunt-Väisälä frequency, and thus the qualitative stability analysis remains unchanged. These trends are robust across all EOSs considered and both tabulated and hybrid approaches. While the presence of spiral arms at early times amplifies the relative error, this discrepancy remains substantial even at later times, where the spiral features have dissipated but the error still reaches levels of $\sim 85\%$ in extended outer regions. The use of Eq.~\eqref{eq:dimiG} is therefore essential to obtain a more accurate characterization of convective stability throughout the remnant. Moreover, the spatial profile of the relative error provides a direct and quantitative diagnostic to identify the regions where the assumption of hydrostatic equilibrium holds and where it fails.

\begin{figure*}
\centering
\includegraphics[width=2.0\columnwidth]{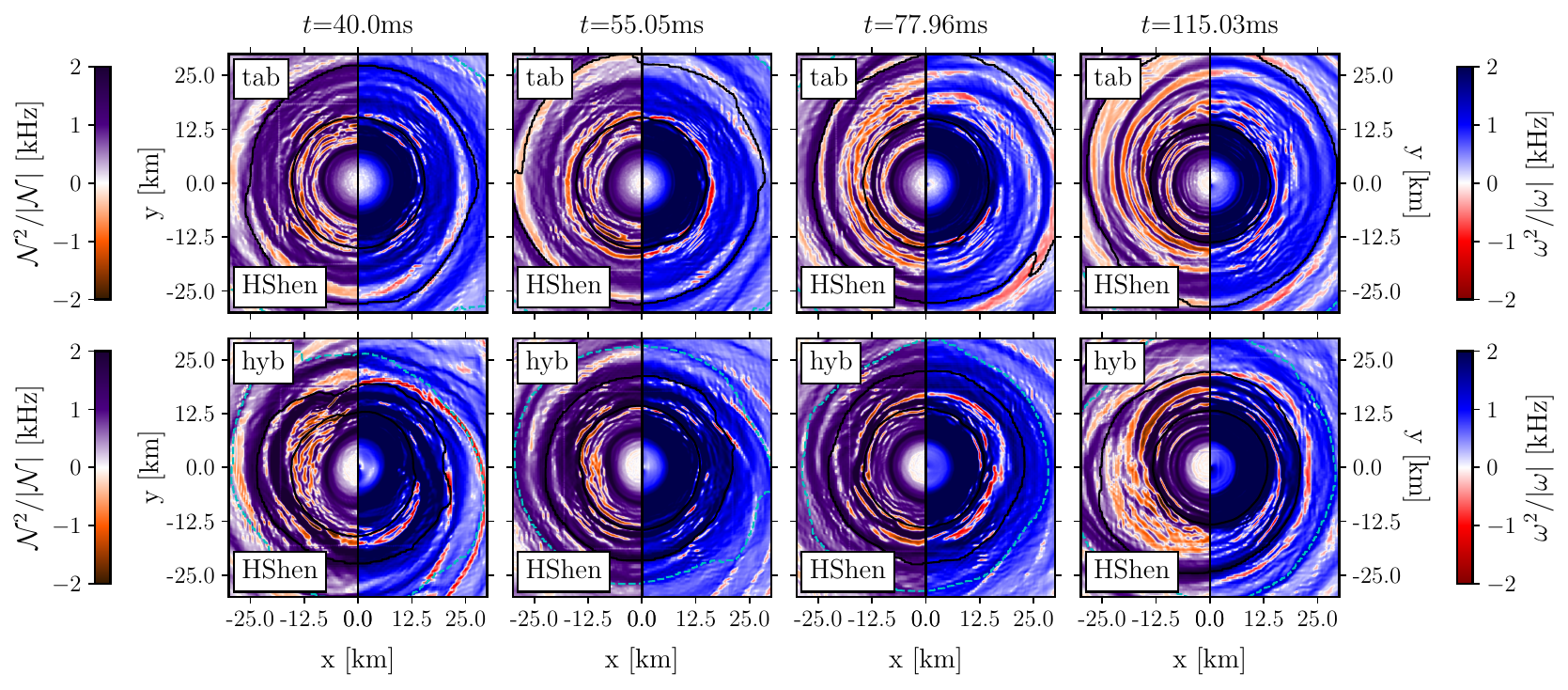}
\caption{Equatorial-plane distribution of the Brunt–Väisälä frequency $\mathcal{N}^2$ (left half of each plot) and the rotationally corrected stability criterion $\omega^2$ (right half of each plot), at four post-merger times for the {\tt HShen} EOS. The tabulated representation is shown on the top row and the hybrid one on the bottom row.  Black and cyan lines have the same meaning as in Fig.~\ref{fig:snapshots}.}
\label{fig:epic_all}
\end{figure*}

Building on this, the sign of $\mathcal{N}^2$ itself provides a direct indication of local stability: a positive value indicates stability against convection, with buoyancy serving as a restoring force. The Ledoux discriminant, $B_r$, also measures stability, particularly regarding thermal and compositional gradients~\cite{DePietri:2018tpx,DePietri2020}. This is known as the "Ledoux criterion": if $B_r < 0$, the region is stable; if $B_r > 0$, it is unstable under convection. The sign of the Brunt-Väisälä frequency is, however, a more general and robust criterion. Unlike the Ledoux discriminant, which assumes orientation of the gravitational field (typically radial), the Brunt-Väisälä frequency remains well-defined even in the strongly deformed and non-spherical geometries found in post-merger remnants. This makes it especially suitable to analyze the results of NR simulations, where convection may develop along complex trajectories. Furthermore, $\mathcal{N}^2$ naturally arises from the linearized equations of motion for perturbed fluid elements, providing a physically motivated stability criterion that can be directly linked to dynamical oscillations and wave propagation within the star. Moreover, the magnitude of $\mathcal{N}^2$ provides insight into the characteristic frequency of local oscillations, allowing a direct comparison with GW signals.

The analysis of the spatial structure of the Brunt–Väisälä frequency at late post-merger times, computed using Eq.~\eqref{eq:dimiG}, reveals that, for all models, the core regions of the HMNS are typically convectively stable, with $\mathcal{N}^2 > 0$. These regions coincide with areas of high thermal energy, often forming a hot central or toroidal structure. Surrounding these cores, extended regions with negative $\mathcal{N}^2$ appear, signaling unstable convection. Over time, thermal energy spreads outward while the shape and strength of the convective zones evolve. Quantitative differences, however, emerge among EOS: {\tt DD2} yields the closest agreement between tabulated and hybrid models, {\tt HShen} and {\tt LS220} both favor instability in the tabulated case, and {\tt SLy4} in the hybrid one. Our results not only confirm the findings  
first reported by~\cite{DePietri:2018tpx,DePietri2020} for purely hybrid EOS representations but they also give further support for the formation of fully developed convectively unstable rings in HMNS remnants. Those convective regions drive the excitation of inertial modes, amplyfing their oscillations and contributing to the GW spectra of late-time BNS merger remnants. 
This analysis underscores the importance of EOS-dependent thermal structure in shaping the post-merger dynamics and highlights the diagnostic power of the Brunt–Väisälä frequency.

\subsubsection{Rotational effects and the epicyclic frequency}
\label{subsec:epicyclic}

A convective stability analysis based solely on the Brunt–Väisälä frequency provides valuable insights into the thermal and compositional structure of the remnant. However, it neglects the influence of rotation, which plays a dominant role in the post-merger dynamics. As discussed in~\cite{Tassoul:2000}, a more complete picture of the local stability of a rotating fluid in Newtonian gravity can be obtained by including the effects of rotation, the so called \textit{Solberg-Høiland} criteria (I and II) for axisymmetric perturbations. In this criteria the \textit{epicyclic frequency}, $\mathcal{K}$, plays an essential role, specially when considering the fluid in the equatorial plane. Combining the Brunt–Väisälä frequency and the epicyclic frequency into a generalized  criterion is particularly relevant in regions where strong rotation coexists with thermal gradients, potentially altering the onset and growth of convective instabilities. This constitutes the so-called Solberg-Høiland criterion I~\cite{Tassoul:2000}, the only relevant of the two when the analysis is restricted to the equatorial plane. Very recently, this criterion has been extended to general relativity by~\cite{Shibata:2025nfj}, and may allow to distinguish between purely thermally driven axisymmetric instabilities and those affected by rotation. It should be emphasized, however, that no such general criterion for {\it non-axisymmetric} instabilities exists. Therefore, here, we use the Solberg-Høiland criterion I, with the corresponding general relativistic corrections, as a guide for convective instability analysis in the equatorial plane, neglecting possible non-axisymmetric effects.

\begin{figure*}
\centering
\includegraphics[width=2.\columnwidth]{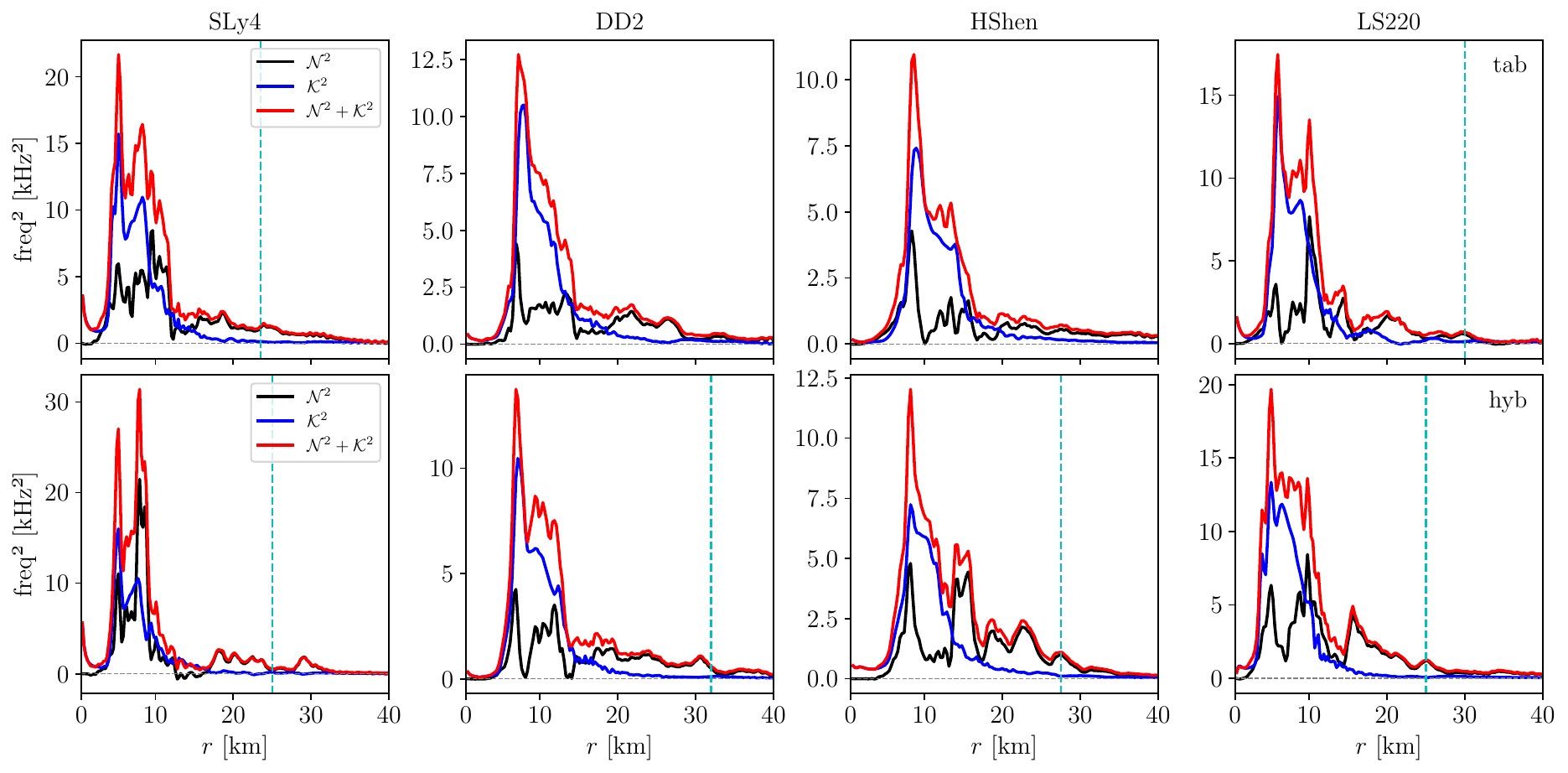}
\caption{Angle-averaged radial profiles of the squared Brunt–Väisälä frequency $\mathcal{N}^2$ (black), epicyclic frequency $\mathcal{K}^2$ (blue), and their sum $\omega^2 = \mathcal{N}^2 + \mathcal{K}^2$ (red) in the equatorial plane. Each column corresponds to a different EOS and each row to a different representation of thermal effects, with tabulated EOSs shown on the top row and the hybrid counterparts on the bottom. All plots refer to $t-t_{\rm merg} = 60\,\rm ms$. The cyan vertical dashed line indicates the approximate coordinate radius of the bulk or the HMNS.}
\label{fig:epic}
\end{figure*}

The epicyclic frequency $\mathcal{K}$ quantifies the natural frequency of radial oscillations of a fluid element in a rotating configuration, such as a HMNS, as it moves along a circular orbit. Physically, it reflects the competition between the inward gravitational pull and the outward centrifugal force experienced by a displaced fluid element. 
Mathematically, $\mathcal{K}$ is defined as the second derivative of the \textit{effective potential}, $V_{\rm eff}$, evaluated at the equilibrium radius $r_0$,
\begin{equation}
    \mathcal{K}^2 = \frac{d^2 V_{\text{eff}}}{dr^2}\bigg|_{r = r_0}\ .
\end{equation}

The effective potential in this context combines gravitational and centrifugal contributions and defines the radial force balance for a fluid element in rotational equilibrium. In Newtonian gravity, and for an angular velocity profile $\Omega(r)$, the epicyclic frequency can be equivalently written as,
\begin{equation}
    \mathcal{K}^2 = 4\Omega^2 + r \frac{d\Omega^2}{dr}\ ,
\end{equation}
showing that $\mathcal{K}$ is not constant but depends explicitly on the radial gradient of the angular velocity. The epicyclic frequency quantifies the rotational restoring force: $\mathcal{K}^2 > 0$ indicates stable oscillatory motion, while $\mathcal{K}^2 < 0$ signifies radial dynamical instability.

Regions with strong negative gradients in $\Omega(r)$ can lead to a reduction of $\mathcal{K}^2$ and potentially trigger rotationally-driven instabilities even when $\mathcal{N}^2$ is positive. Conversely, large positive values of $\mathcal{K}^2$ act as a stabilizing mechanism, suppressing convection. Therefore, a consistent stability criterion in rotating remnants must include both $\mathcal{N}^2$ and $\mathcal{K}^2$, as we now proceed to investigate. Following~\cite{Shibata:2025nfj} we use the relativistic epicyclic frequency, 
\begin{equation}
    \mathcal{K}^2 = \gamma_{rr}^{-1/2}\, \theta_r\, \partial_r \ell \ ,
\label{eq:epicyclic}
\end{equation}
where $\ell = -u_\phi / u_t$ is the specific angular momentum of the fluid. The quantity $\theta_r$ is the radial component of the orientation vector $\theta^i$, which encodes the relativistic correction to the centrifugal force and is defined by~\cite{Shibata:2025nfj},
\begin{equation}
\theta_i = \left( u^t u_t \right)^2 \left[ \frac{1}{W^2 \gamma_{\phi\phi} u_t^2} \partial_i \ell - \partial_i \Omega \right]\ .
\end{equation}
Here, $\Omega = u^\phi / u^t$ is the fluid's angular velocity, $\gamma_{\phi\phi}$ is the azimuthal metric component, and $W$ is the Lorentz factor. 

\begin{figure*}
\begin{center}
\includegraphics[width=2.\columnwidth]{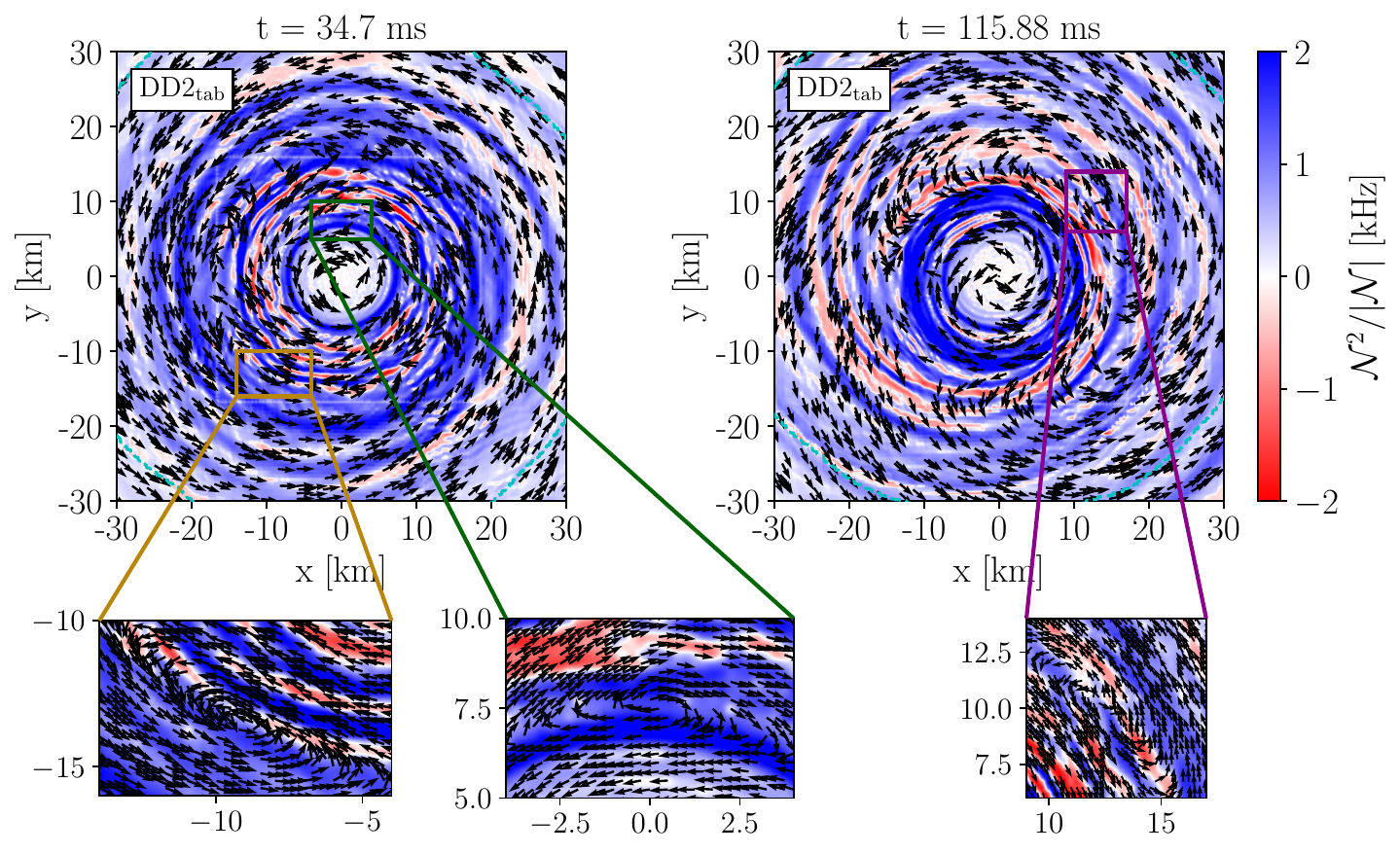}
\caption{Equatorial-plane snapshots of the normalized velocity field (black arrows) overlaid with the Brunt-Väisälä frequency (colormap) for the {\tt DD2} tabulated EOS. The left panel shows an early post-merger time ($t-t_{\rm merg}=34.7$ ms) and the right panel a later time ($t-t_{\rm merg}=115.88$ ms). The cyan dotted line represents the remnant's bulk, and zoomed plots highlight identified convective regions.}
\label{fig:velrot_dd2}
\end{center}
\end{figure*}

The combined effect of the epicyclic frequency and the Brunt–Väisälä frequency, thus, governs convective stability of axisymmetric perturbations, reflecting the interplay between thermal stratification (buoyancy) and rotational support (centrifugal forces). Using the previous relativistic definitions for $\mathcal{N}^2$ and $\mathcal{K}^2$, we can express the Solberg-Høiland stability criterion I in a form similar to its Newtonian counterpart,
\begin{equation}
    \omega^2 \coloneqq \mathcal{N}^2+\mathcal{K}^2 > 0 \ .
\label{eq:criterion}
\end{equation}
This condition implies that both epicyclic and buoyancy oscillations must be simultaneously considered to assess whether a region is stable against axisymmetric perturbations. If $\omega^2 < 0$, the region is dynamically unstable. This criterion is particularly relevant in the study of the internal structure of rotating NSs, where differential rotation, relativistic effects, and compositional gradients play a crucial role in the overall stability~\cite{reisenegger:1992,Paschalidis:2016agf,Shibata:2025nfj}.

As an illustrative example of our EOS sample, Fig.~\ref{fig:epic_all} displays the equatorial-plane distribution of $\mathcal{N}^2$ and $\omega^2$ at four representative post-merger times for the {\tt HShen} EOS, with the tabulated model on the top row and the hybrid one at the bottom. We note that in the computation of the Brunt–Väisälä frequency we use Eq.~\eqref{eq:dimiG} for the definition of the gravitational acceleration. It is worth mentioning that the results using Eq.~\eqref{eq:shibataG}, employed in~\cite{Shibata:2025nfj} assuming hydrostatic approximation, yield values of $\omega^2$ everywhere positive at all times (not shown in Fig.~\ref{fig:epic_all}). This sharply contrasts with the results obtained using Eq.~\eqref{eq:dimiG}. In this case, $\omega^2$ is initially positive in the inner core, reflecting the stabilizing effect of rotation, but unstable rings with $\omega^2 < 0$ soon appear as buoyancy overcomes centrifugal support. This is a general trend observed in all EOS of our sample, and both for tabulated and hybrid representations. Those unstable regions persist at later times, though they are weaker than in the analysis employing $\mathcal{N}^2$ alone (compare right and left halves of each plot). Therefore, assuming hydrostatic equilibrium as in~\cite{Shibata:2025nfj} seems to incorrectly suggest global stability despite the presence of strong thermal gradients and observed convection (see Section~\ref{sub:convectiveregions}).\\

To complement the equatorial-plane analysis of convective stability, we now present an angle-averaged study of the Brunt–Väisälä frequency $\mathcal{N}^2$, the epicyclic frequency $\mathcal{K}^2$, and their sum $\omega^2 = \mathcal{N}^2 + \mathcal{K}^2$. This global radial profiling follows the methodology introduced in~\cite{Shibata:2025nfj}, where the angular average over spherical shells yields a coordinate-independent assessment of stability. Fig.~\ref{fig:epic} shows radial profiles of these quantities for all EOSs at $t-t_{\rm merg}=60\,\rm ms$. While angle-averaged profiles are useful in identifying {\it global} stability trends, they may not accurately reflect the presence of {\it local} instabilities at a given radius. As shown in our previous equatorial-plane analysis, many of the convective structures, especially ring-like features, are not perfectly circular. Instead, they display azimuthal asymmetries and may only partially extend over a given angular sector. When averaging over spherical shells, these asymmetries can be smoothed out, resulting in average values of the relevant frequencies that remain positive even when some angular sectors are locally unstable. This effect is illustrated in Fig.~\ref{fig:epic}, where the angle-averaged values of $\mathcal{N}^2$ and $\omega^2$ appear strictly positive in the inner $\sim 20\,\rm km$ at $t = 60\,\rm ms$ for all models, suggesting stability. However, equatorial-plane distributions clearly show localized negative regions in the same radial range (see the snapshots at $t-t_{\rm merg}=55\,\rm ms$ and 78 ms in Fig.~\ref{fig:epic_all} for the {\tt HShen} EOS). These discrepancies highlight that the average at a given radius may simply reflect the dominance of positive cells over negative ones in the angular integration. As a result, ring-like instabilities that are partially developed or non-uniformly distributed in $\phi$ may manifest as local minima in the radial average of $\mathcal{N}^2$ or $\omega^2$, without ever becoming negative. Therefore, the conclusions from an angle-averaged approach must be interpreted with caution. 

\subsection{Visualizing convective regions in the HMNS}
\label{sub:convectiveregions}

To further identify convective structures in the post-merger remnant, it is useful to analyze the equatorial velocity field. This involves normalizing the velocity vector field by its local magnitude to focus on direction, calculating the rotation profile $\Omega(r)$ through angle-averaging (see Section~\ref{sub:rotation}), and then subtracting constant angular velocities from the vector field. This decomposition isolates fluid motion faster or slower than the reference frame.
Subtracting a constant angular velocity $\Omega_0$ reveals three regions: an inner core and outer envelope where $\Omega(r) < \Omega_0$, separated by an intermediate region where $\Omega(r) > \Omega_0$. At radii where $\Omega(r) = \Omega_0$, the fluid transitions through zero angular velocity, forming rings in the equatorial plane. Spiral patterns and other convective features preferentially appear near these rings, suggesting that subtracting the background rotation exposes underlying residual fluid dynamics like buoyancy or vorticity. 

To systematically explore the rotational structure and its connection to convection we apply this procedure for four different values of $\Omega_0$, covering a broad range of the angular velocity profile. This analysis was performed for both hybrid and tabulated EOS approaches. By subtracting each constant $\Omega_0$ value from the full velocity field, we are able to isolate differential motion and identify convective features more clearly. We find that in both the hybrid and tabulated models, convective regions consistently form and persist throughout the entire simulation, remaining visible even at late times (beyond $120$ ms after merger). Fig.~\ref{fig:velrot_dd2} illustrates these results for a representative case using the tabulated {\tt DD2} EOS. We show the normalized velocity field with black arrows overlaid with the Brunt–Väisälä frequency $\mathcal{N}^2$. From the visual analysis of the resulting vortex structures, we observe the following trends:
\begin{itemize}
    \item A large number of vortex-like features appear within the first millisecond after merger, especially in regions near or within unstable rings (where $\mathcal{N}^2 < 0$).
    
    \item As the system evolves, the number of vortices decreases. Typically, we observe one prominent vortex structure in the inner core and other one or two vortices in the outer layers.
    
    \item Vortices are found either inside or just outside the unstable rings defined by the intersections of $\Omega(r)$ and $\Omega_0$.
    
    \item When an unstable region propagates through the fluid, it may displace, distort, or even destroy existing vortex structures.
    
    \item After the unstable ring passes, new vortices may form in its wake, indicating a coupling between transient convective layers and rotational shear.
\end{itemize}

\noindent The general behavior of these convective regions is similar for stiff and soft EOSs, but there are some variations between the tabulated and hybrid approaches. For {\tt DD2}, differences between the two approaches are minimal. However, for {\tt HShen}, the hybrid EOS exhibits more convective regions compared to the tabulated one. Finally, the tabulated {\tt SLy4} EOS shows only one or two more convective regions than the hybrid counterpart. 

\subsection{Rotational profiles}
\label{sub:rotation}

\begin{figure}
\begin{center}
\includegraphics[width=1.\columnwidth]{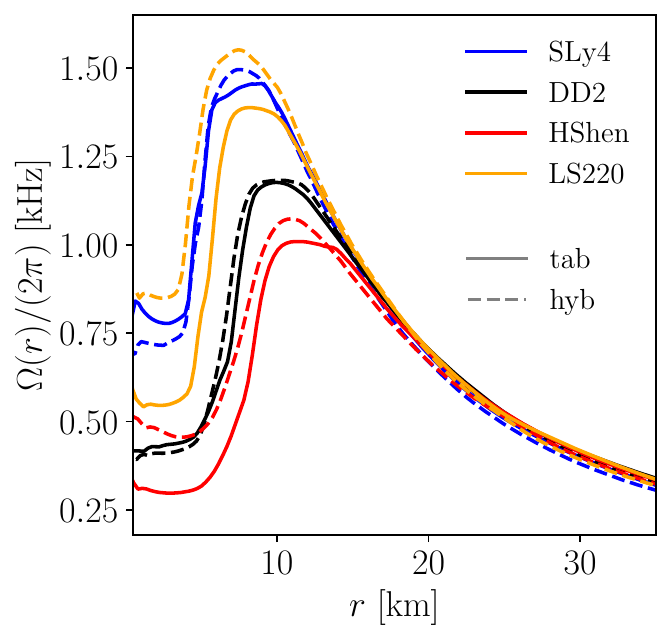}
\includegraphics[width=1.\columnwidth]{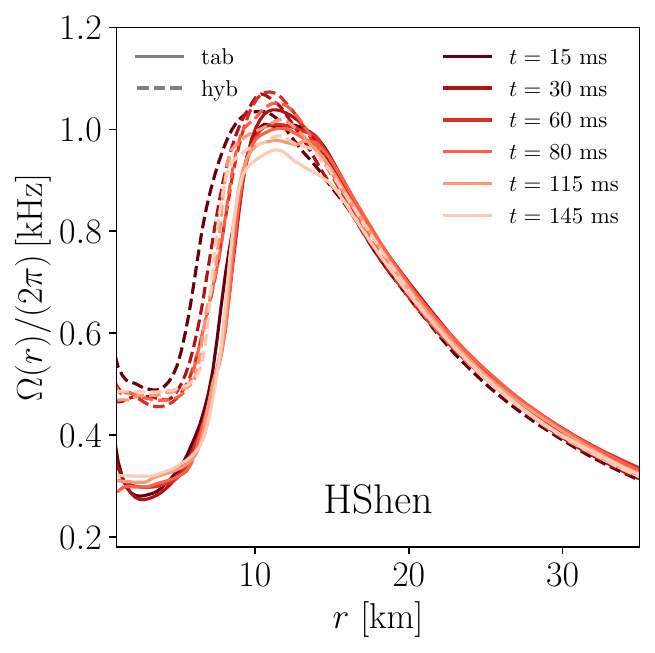}
\caption{Top: Rotational profiles $\Omega(r)/2\pi$ versus coordinate radius $r$ for all EOSs at $t-t_{\rm merg}=60\,\rm ms$.  Bottom: Snapshots of the rotational profile for the {\tt HShen} EOS at various post-merger times. In both panels solid lines correspond to tabulated EOSs and dashed lines to the hybrid counterparts. }
\label{fig:rotationlaw}
\end{center}
\end{figure}

The remnant of a BNS merger is typically supported against gravitational collapse by differential rotation. Its angular velocity $\Omega$ is not constant and varies with the radial distance from the core, forming a differentially rotating structure~\cite{Baumgarte_2003}. A commonly used empirical expression for the angular velocity profile in a HMNS is given by the expression in the Newtonian limit~\cite{Uryu:2016dqr} $\Omega(r) = \Omega_c/(1 + (r/r_0)^n)$, where $\Omega_c$ is the central angular velocity, $r_0$ is a characteristic radius at which the angular velocity drops significantly, and $n$ determines the steepness of the rotational gradient. This profile captures the essential features of differential rotation, with $\Omega(r)$ decreasing with the distance from the center. In numerical simulations, the angular velocity can also be computed from the tangential velocity $v_\phi$ and the cylindrical radial coordinate $r$ using the relation
\begin{equation}
\Omega = \frac{v_\phi}{r}\ .
\end{equation}
The rotational profile is essential for the stability of the remnant and influences the excitation of oscillation modes. Sharp gradients in $\Omega(r)$ lead to shear instabilities and dynamic processes that impact the remnant's long-term evolution. 

The top panel of Fig.~\ref{fig:rotationlaw} displays the rotational profile $\Omega(r)/2\pi$ for all EOSs, both tabulated (solid lines) and hybrid (dashed lines), at $t-t_{\rm merg}=60\,\rm ms$. For most EOS, the rotational profile reaches higher frequencies in the hybrid cases compared to the tabulated ones, primarily at the center of the remnant. The only exception is the {\tt SLy4} EOS, where the tabulated model rotates faster than the hybrid one at the center. This is also apparent in the bottom panel of Fig.~\ref{fig:rotationlaw}, which shows the rotational profile computed at various times after the merger for the {\tt HShen} EOS. This figure shows that remnants evolved with the hybrid EOS rotate faster than those with tabulated EOS. The differences in the rotational profiles are ultimately attributed to how temperature and thermal energy effects are included in the modelling. The more accurate treatment of those effects in tabulated EOSs leads, in the absence of other effects such as neutrino cooling and magnetic dissipation, to less rapidly rotating remnants than in the case of hybrid EOSs.

\section{Conclusions}
\label{Conclusions}

We have discussed long-term, numerical-relativity simulations of binary neutron star mergers modeled using both, fully tabulated, finite-temperature, equations of state and their corresponding {\it hybrid} representations. The latter combine a cold piecewise-polytropic part and a thermal ideal-gas-like part. The motivation of this study has been to improve our understanding of how the treatment of finite-temperature effects in the equation of state influences the dynamics of post-merger remnants. In particular, we have attempted to address whether the late-time excitation of inertial modes in post-merger remnants, previously reported in studies based solely on hybrid EOS~\cite{DePietri:2018tpx,DePietri2020}, is still supported when using tabulated, finite-temperature EOS which account for thermal effects in a more consistent way. Indeed, our results show that such inertial modes do persist, even though the presence of convectively unstable regions in the remnant, directly responsible for the excitation of the modes, is somewhat less apparent in simulations based on tabulated EOS than on their hybrid counterparts.

Differences between the two approaches are already present, to some extent, in the initial data. We have compared initial data built with tabulated EOS models and with $7-$ and $10-$pieces hybrid representations. This has highlighted significant differences that can affect the realism of the simulations. While the hybrid EOS representation is more computationally efficient, it may fail to capture the complexities of NS matter, especially with a limited number of pieces. Attempting to increase the number of pieces to better approximate the tabulated EOS poses challenges, including higher computational demands and practical limitations. Critically, this approach does not guarantee convergence to the tabulated EOS. This study has shown that even with $10$ pieces, discrepancies persist between the initial conditions obtained from the hybrid representation and those derived from tabulated EOS data, influencing simulation results.  

We have investigated the spectra of the post-merger GW signals and how these are affected by the treatment of thermal effects in the two EOS representations. This has allowed us to derive new quasi-universal relations for the main post-merger GW frequencies, associated with oscillation modes of the remnant. In particular, the effect of the EOS on the frequency and amplitude of post-merger GW signals has been studied using the fundamental quadrupolar $f_2$ mode and the inertial modes. The $f_{2,i}$ mode, prevalent in the early post-merger phase, does not differ much between hybrid and tabulated EOSs, while the $f_2$ and inertial modes appear at different times and their frequencies are shifted between the two EOS approaches. Our simulations also highlight distinct differences in the GW frequency evolution related to the thermal modeling of the EOS, demonstrating that deviations from established quasi-universal relations become significant at late post-merger phases. Combining the results from~\cite{Rezzolla2016,Topolski:2023ojc,Rivieccio_2024} with our results, we have observed that the quasi-universal relation previously reported for the instantaneous (peak) frequency at merger remains essentially unaltered. However, significant deviations between our empirical fits and those reported in~\cite{Rezzolla2016,Topolski:2023ojc} are observed at later post-merger times. Deriving new empirical fits is particularly relevant for conducting inferences on NS properties using the post-merger GW signal and highlights the importance of incorporating thermal effects in the EOS as accurately as possible.

Our simulations have also illustrated how the inclusion of finite-temperature effects in the EOS influence the fate of the remnant. Simulations employing tabulated EOSs are less prone to gravitational collapse due to their improved handling of thermal gradients and pressure support, while hybrid EOS models tend to produce more compact remnants, leading to their premature collapse. This effect is particularly evident for soft EOS such as {\tt LS220}, where the tabulated model can sustain stability through thermal pressure and differential rotation, whereas the hybrid model collapses to a black hole due to insufficient thermal support. 

The convective stability of the remnant has been examined  in detail, employing both the Ledoux criterion, a necessary condition for the development of instabilities, and the Solberg-Høiland criterion I~\cite{Tassoul:2000}, a generalized criterion based of a combined analysis of the Brunt–Väisälä frequency and the epicyclic frequency, valid only for axisymmetric perturbations. This analysis has shown that fully tabulated EOSs predict distinct convective patterns at late post-merger times, especially in the bulk of the remnants. Convection, however, is somewhat more apparent in the simulations employing hybrid EOSs. These differences directly affect the evolution of inertial modes, modifying their growth and sustained excitation compared to hybrid models. Our stability analysis suggests that inertial modes are influenced by temperature and rotation variations, which are represented more accurately by tabulated EOS models. We have argued that even though an angle-averaged analysis based on the convective stability criterion seems to indicate that the remnant is {\it globally} stable (see also~\cite{Shibata:2025nfj}), the significant deviation from spherical symmetry of the post-merger remnant still allows for the development of {\it local} convectively unstable regions (and convection patterns in the velocity fields). Those local convective patterns, while substantially different between tabulated and hybrid EOSs, remain visible throughout the entire simulated time, triggering the excitation of inertial modes. Their frequencies are smaller than those attained by the fundamental quadrupolar mode and are potentially within reach of third-generation GW detectors. 

Another outcome of this study has been the assessment of thermal effects on the rotation profiles of the remnants. We have found that tabulated EOSs produce remnants with less extreme rotational gradients, which are critical for stability, while hybrid EOSs often lead to sharper gradients that can induce shear instabilities. The rotational profile directly impacts the GW emission by affecting mode frequencies and amplitudes, as well as the overall detectability of the signal. Accurate temperature-modeling in the EOS is therefore essential for understanding the post-merger rotation profile and its implications for remnant stability and GW emission. 

Ultimately, this study stresses the importance of how to incorporate thermal effects in long-term simulations of BNS merger remnants. We have shown that different ways of doing that, either through fully tabulated EOS derived from nuclear physics or using more simplistic hybrid counterparts, can lead to substantial differences in simulation outcomes, as seen in the varied thermal profiles in the bulk produced by either approach. Accurate temperature modeling is particularly relevant for interpreting the thermal evolution of the remnants, as it quantitatively affects their stability against gravitational collapse, their dynamics, the excitation of oscillation modes, and the associated GW spectra. The findings reported in this paper, in particular the fact that inertial modes can be excited and may become a promising way to infer thermal and rotational properties of neutron stars through GW asteroseismology,  should be complemented with further simulations accounting for additional physics, such as magnetic field amplification (and their associated instabilities) and neutrino cooling. Those simulations will be the subject of future work.

\begin{acknowledgments}
The authors thank Zacharias Etienne and Leonardo Werneck for useful comments. The authors thanks also Roberto De Pietri, Alessandra Feo, Frank Löffler, and Francesco Maione for developing the {\sc pycactus} python post-processing package for analysing output of {\tt Eistein Toolkit} simulations. This work has been supported by the Generalitat Valenciana through the grants CIDEGENT/2021/046
and Prometeo CIPROM/2022/49; by the Spanish Agencia Estatal de  Investigaci\'on through the grants PRE2019-087617,  PID2021-125485NB-C21, PID2023-147112NB-C22 and PID2024-159689NB-C21 funded by 
MCIN/AEI/10.13039/501100011033 and ERDF A way of making Europe;
through grant CNS2022-135529 within the “European Union NextGenerationEU/PRTR” and through the “Unit of Excellence María de Maeztu 2025-2031” Award CEX2024-001451-M to
the Institute of Cosmos Sciences; and by the 
Generalitat de Catalunya through Grant 2021SGR01069. 
Further support has been provided by the EU's Horizon 2020 Research and Innovation (RISE) programme H2020-MSCA-RISE-2017 (FunFiCO-777740), by  the  EU  Staff  Exchange  (SE)  programme HORIZON-MSCA-2021-SE-01 (NewFunFiCO-101086251), and by MCIN and Generalitat Valenciana with funding from NextGenerationEU (PRTR-C17.I1, Grant ASFAE/2022/003). The authors acknowledge the computational resources and technical support of the Spanish Supercomputing Network through the use of MareNostrum at the Barcelona Supercomputing Center (AECT-2023-1-0006 and AECT-2023-1-0023) where the simulations were performed and the resources from the Gravitational Wave Open Science Center, a service of the LIGO Laboratory, the LIGO Scientific Collaboration and the Virgo Collaboration. M.P. acknowledges the financial support from the PNRR MUR project ECS-00000033-ECOSISTER. This work has used the following open-source packages: \textsc{NumPy}~\cite{harris:2020}, \textsc{SciPy}~\cite{scipy:2020}, \textsc{Matplotlib}~\cite{Hunter:2007}, \textsc{Lorene}~\cite{Lorene}, \textsc{EinsteinToolkit}~\cite{EinsteinToolkit:2024_05}, \textsc{IllinoisGRMHD}~\cite{Etienne:2015cea,Noble:2005gf}, \textsc{Kuibit}~\cite{kuibit}, and \textsc{pycactus}~\cite{pycactus}.
\end{acknowledgments}


\appendix

\section{Temperature definition: tabulated vs hybrid EOSs}
\label{app:Temperature}

In this Appendix, we provide a brief discussion of the calculation of temperatures in both the hybrid and the tabulated approaches. In particular, we  discuss whether the prescription introduced in Ref.~\cite{DePietri2020} and explained below provides a good estimate for the temperature for hybrid EOS descriptions. While a system in thermodynamical equilibrium has a well defined temperature in principle, in practice it may be difficult to assign a temperature to hydrodynamically evolving matter cells. In particular, depending on the treatment of the underlying matter, the temperature may be obtained in different ways in numerical simulations of BNS mergers. 

In hybrid models, one typically separates the zero-temperature and thermal contributions
of the energy per particle,
\begin{align}
e(\rho,T) = e_\text{cold}(\rho) + e_\text{th}(\rho,T)\ ,
\end{align}
and the pressure,
\begin{align}
P(\rho,T) = P_\text{cold}(n) + \left( \Gamma_\text{th} - 1 \right)\, \rho\,  e_\text{th}(\rho,T)\ .
\label{eq:hyb_pressure}
\end{align}
As discussed in detail in Appendix C of Ref.~\cite{DePietri2020}, thermodynamical consistency allows one to define an approximation
to the temperature of the system from the expression,
\begin{align}
T_\text{th} = \left( \Gamma_\text{th} - 1 \right)\, e_\text{th}(\rho,T)\,,
\label{eq:thermal_temperature}
\end{align}
in hybrid EOS models. In this expression the thermal index is given by,
\begin{align}
\Gamma_\text{th} = 1 + \frac{P(\rho,T) - P_\text{cold}(\rho)}{\rho \left[ e(\rho,T) - e_\text{cold}(\rho)  \right]}\ .
\label{eq:thermal_index}
\end{align}

\begin{figure}
\begin{center}
\includegraphics[width=1.\columnwidth]{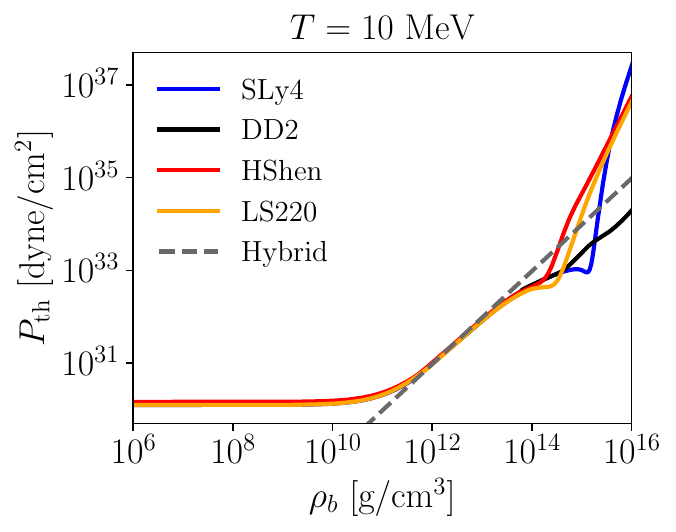}
\caption{Thermal pressure as a function of matter density for all EOSs at a fixed temperature of $10\, \rm MeV$ and electron fraction $Y_e=0.065$, representative of the postmerger phase. The grey dashed line represents the analytic thermal pressure for the hybrid case at this temperature.}
\label{fig:pthermal}
\end{center}
\end{figure}

In some ideal situations, like the ideal gas or non-interacting degenerate matter, $\Gamma_\text{th}$ is a constant, independent of density or temperature. In particular, for an ideal gas, one has $P_0=0$ and 
\begin{equation}
    P_{\rm th}(\rho,T) = T\, \rho\ .
\label{eq:pth_hybrid}
\end{equation}
One also obtains the same result by employing the thermodynamically consistent prescription of Eq.~\eqref{eq:hyb_pressure} and Eq.~\eqref{eq:thermal_temperature}. 

In a tabulated EOS, however, the thermal pressure incorporates interaction effects, clustering and other correlations. As a consequence, it is not necessarily a simple function of density and temperature. 
We compare the thermal pressure from the analytical formulation with that from the tabulated EOS at a given temperature $T$ and electron fraction $Y_e$ in Fig.~\ref{fig:pthermal}. 
We pick the value $Y_e\simeq 0.065$ for this comparison. This is the average value of $Y_e$ in and around the remnant along the equatorial plane $20\,\rm ms$ after merger (when the remnant begins to relax resulting in a more circular surface of the bulk). We choose a temperature of $T = 10\,\rm MeV$, close to the average temperature during the evolution as discussed in the context of Fig.~\ref{fig:bulktemp}. 

Figure~\ref{fig:pthermal} shows the thermal pressure $P_{\rm th}$ as a function of $\rho$ for each EOS model (colored lines) and for the definition in Eq.~\eqref{eq:pth_hybrid} (dashed gray line). At low densities, the tabulated EOSs  thermal pressure remains is constant, with no distinction among the EOSs. The thermal pressure at low temperatures is dominated by photons, and it is strongly temperature dependent but density independent. At larger densities and all the way up to $\approx 10^{13}\, \rm g/cm^3$, all the tabulated approaches agree. Moreover, between $10^{12}\, \rm g/cm^3$ and $10^{13}\, \rm g/cm^3$, the tabulated approaches resemble
the ideal gas behaviour of Eq.~\eqref{eq:pth_hybrid}, shown with a dashed grey line. The ideal gas and the tabulated EOS diverge   significantly at densities above saturation, $\approx 10^{14}\, \rm g/cm^3$. 

In this high-density region, in fact, every EOS predicts a somewhat different thermal pressure. 
The {\tt DD2} EOS appears to be the closest to the hybrid behaviour. At high densities, it deviates from Eq.~\eqref{eq:pth_hybrid} by providing a somewhat smaller thermal pressure with a very similar density dependence. This may explain why the {\tt DD2} case consistently shows the smallest differences between tabulated and hybrid approaches in the results shown across this paper. In contrast, all the other EOS predict thermal pressures at large densities that are substantially larger and with a steeper density dependence than the hybrid approach. EOS-dependent effects associated to degeneracy, nuclear interactions and correlations such as clustering dominate the thermal behaviour of matter in this region~\cite{carbone2019,Rivieccio:2025zcw}. 

In a numerical simulation, the differences in thermal behaviour arise from at least 3 different sources. First, an explicit EOS table propagation has a different thermal behaviour, as indicated in Fig.~\ref{fig:pthermal}. Second, these differences propagate over time, providing different histories and hence different simulation outcomes, as discussed earlier in this paper. Third, the local temperature is well defined in the EOS approach, whereas in a hybrid model one relies on Eq.~\eqref{eq:thermal_temperature} to extract a "local temperature" $T_\text{th}$ for fluid elements. 
We now analyze whether this approach in the hybrid setting provides a good estimate of the actual temperature of the system in two different physical limits, the ideal classical gas  and the degenerate gas. To this end, we employ microphysics simulations that compute $T_\text{th}$ for pure neutron matter and compare them to the analytical expressions we obtain in limiting cases.

\subsection{Ideal classical gas}
At low densities and high temperature, the ideal classical approximation should hold. Figure~\ref{fig:pthermal} indeed suggests that all EOS have an "ideal region window" as a function of density for a given temperature.  We now consider a single-component gas of non-relativistic neutrons for simplicity, although the conclusions do not change for the relativistic single-component case. In this approximation, the zero-temperature components of the energy and the pressure are 
negligible, $e_\text{cold}(\rho)=P_\text{cold}(\rho)=0$, compared to their thermal 
counterparts, so that
$e(\rho,T) \approx e_\text{th}(\rho,T) = 3/2\, T$
and
$P(\rho,T) \approx P_\text{th}(\rho,T)$.
In this ideal setting, the thermal index of Eq.~\eqref{eq:thermal_index} is simply $\Gamma_\text{th}=5/3$, regardless of the temperature and density, and the estimate of Eq.\eqref{eq:thermal_temperature} provides:
\begin{align}
T_\text{th} = \left( \frac{5}{3} - 1 \right) \frac{3}{2} T = T\ .
\end{align}
In other words, we expect that in the low-density, high-temperature regime, where matter behaves like an ideal gas, $T_\text{th}$ should provide a good approximation to the temperature of the system~\cite{Rivieccio:2025zcw}.  This result should be relatively general, in that interactions do not play a qualitative role in the low-density regime, and having more than one component only introduces a mild density dependence in the thermal  index~\cite{NadalTFG,Rivieccio:2025zcw}.

\subsection{Ideal degenerate gas}
Let us now consider a single-component, non-relativistic, degenerate neutron system. This approximation should be valid whenever the ratio of the temperature to the Fermi energy, $T/\varepsilon_F$, is small.
The Fermi energy is defined as $\varepsilon_F = (\hbar^2 k_F^2)/(2 m_n)$, with $m_n$ the neutron mass and the Fermi momentum $k_F=(3\, \pi^2\, n)^{1/3}$ and $n$ is the number density, such that $\rho=mn$. In the degenerate regime, employing the Sommerfeld expansion~\cite{Huang,weiss_cox_2004}, one finds the energy per particle
\begin{align}
e(\rho,T) = \frac{3}{5} \varepsilon_F \left[  1 + \frac{5 \pi^2}{12} \left( \frac{T}{\varepsilon_F}\right)^2 \right]\ ,
\end{align}
and the pressure,
\begin{align}
P(\rho,T) = \frac{2}{5} \varepsilon_F\,n \left[  1 + \frac{5 \pi^2}{12} \left( \frac{T}{\varepsilon_F}\right)^2 \right]\  .
\end{align}
The thermal component to the energy reads
\begin{align}
e_\text{th} (\rho,T) =  \frac{\pi^2}{4} \frac{T^2}{\varepsilon_F}\ ,
\label{eq:e_thermal_deg}
\end{align}
whereas the thermal component of the pressure is
\begin{align}
P_\text{th} (\rho,T) =  \frac{\pi^2}{6} \frac{T^2}{\varepsilon_F}n\ .
\label{eq:p_thermal_deg}
\end{align}
Note that even though $e_\text{th} (\rho,T)$ and $P_\text{th} (\rho,T)$ are density and temperature dependent, the thermal index
is still a constant in this case, $\Gamma_\text{th}=5/3$. Employing the Sommerfeld expansion in Eq.~\eqref{eq:thermal_temperature}, however, we find that the temperature estimate is not proportional to the temperature of the system anymore. Instead, it is given by the extession
\begin{align}
T_\text{th} = \left( \frac{5}{3} - 1 \right)\, e_\text{th} (\rho,T)  =  \frac{\pi^2}{6} \frac{T^2}{\varepsilon_F}\ .
\label{eq:thermal_degenerate}
\end{align}
In other words, in the degenerate regime, $T_\text{th}$ is not a good proxy for the temperature. In fact, for a constant temperature $T$, $T_\text{th}$ is a decreasing function of Fermi energy and, hence, density, that should vanish at sufficiently high densities.

\begin{figure}[t!]
\centering
\includegraphics[width=1.\linewidth]{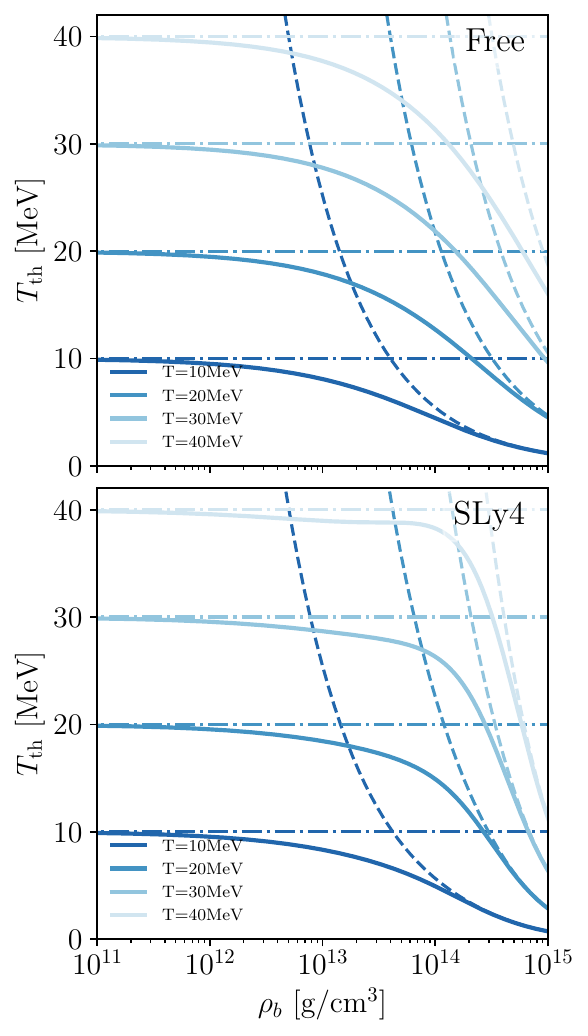}
\caption{Top panel: thermal temperature estimate $T_\text{th}$ of Eq.~\eqref{eq:thermal_temperature} as a function of density for $4$ different temperatures (solid lines). The horizontal dot-dashed lines correspond to the $4$ actual temperatures. The dashed line, matching the numerical results at high densities, corresponds to the ideal degenerate gas estimate of Eq.~\eqref{eq:thermal_degenerate}. Bottom panel: the same for a gas of neutrons interacting with the {\tt SLy4} Skyrme parametrization. The degenerate estimate is modified to account for interactions employing Eq.~\eqref{eq:thermal_degenerate_Sk}.}
\label{fig:Appendix_thermal}
\end{figure}

We explore the validity of these two approximations in the top panel of Fig.~\ref{fig:Appendix_thermal}. We employ a full numerical solution of the thermodynamical properties of a gas of non-relativistic neutrons to compute $e_\text{th}$ exactly, and employ Eq.~\eqref{eq:thermal_temperature} to estimate the temperature of the system. The simulations cover a range of densities between $10^{11}\,\rm g/cm^3$ and $1\times10^{15}\,\rm g/cm^3$, and have been performed at $T=(10,20,30,40)\,\rm MeV$. We stress that the neutron matter approximatione employed here is an idealised scenario and should not be taken as a representative of the dense matter inside a BNS. 
We find that the thermal temperature estimate, based on Eq.~\eqref{eq:thermal_temperature} and shown in solid lines, is only a valid approximation to the true temperature at asymptotically low densities, well below $\rho_b \approx 10^{12}\,\rm g/cm^3$. Above this density, the estimate based on $T_\text{th}$ underestimates the real temperature substantially. We also find that, for sufficiently high densities (e.g.~above $10^{14}\,\rm g/cm^3$ at $T=10\,\rm MeV$), the calculation of $T_\text{th}$ employing Eq.~\eqref{eq:thermal_temperature} with $e_\text{th}$ taken from a full numerical solution agrees with the degenerate gas estimate of Eq.~\eqref{eq:thermal_degenerate}. In particular, in agreement, with this equation, we find that $T_\text{th}$ tends to zero asymptotically as the density increases.

\subsection{Interacting degenerate gas}
We do not expect interaction effects to change qualitatively these conclusions at the mean-field level~\cite{constantinou2015,carbone2019}. For the same degenerate conditions discussed earlier, there are two types of modifications that come into play. First, the momentum-dependence of the interaction modifies the mass of neutrons in the medium, which acquire a density-and possibly temperature-dependent effective mass, $m_n^*$~\cite{constantinou2015}. The Fermi energies of Eqs.~\eqref{eq:e_thermal_deg} and \eqref{eq:p_thermal_deg} are thus renormalized, $\varepsilon_F \to \varepsilon_F^* =  (\hbar^2\, k_F^2)/(2\, m_n^*)$. In addition, the thermal pressure acquires an additional term associated to the density dependence of $m_n^*$~\cite{constantinou2015}, and the thermal index becomes
\begin{align}
\Gamma_\text{th}^* =  \frac{5}{3}  - \frac{\rho}{m_n^*} \frac{\partial m_n^*}{ \partial \rho}\ .
\label{eq:gamma_thermal_deg_int}
\end{align}
Typically, this index is somewhat  density dependent and mildly temperature dependent, although specific results depend on the underlying nuclear physics assumptions~\cite{constantinou2015,carbone2019,Keller2023}. Regardless of the model, however, one expects that interaction effects should play a role only at the quantitative, but not the qualitative, level.

The bottom panel of Fig.~\ref{fig:Appendix_thermal} shows the results of a numerical simulation of $T_\text{th}$ for the neutron system interacting with the {\tt SLy4} density functional~\cite{Chabanat97,Chabanat97un} (solid lines). Just as in the (free) non-interacting case (top panel), we find that $T_\text{th}$ underestimates the real temperature of the system across a wide density regime, particularly above $10^{12}\,\rm g/cm^3$. Compared to the top panel, we find that, at temperatures above $\approx 20\, \rm MeV$, $T_\text{th}$ stays a bit closer to the real temperature of the system at low densities. However, above a certain point, a sharp transition occurs and $T_\text{th}$ decreases drastically as a function of density.

At relatively large densities, we can again employ the Sommerfeld expansion to obtain a closed expression for $T_\text{th}$ in the interacting case. Using the thermal index of Eq.~\eqref{eq:gamma_thermal_deg_int}, the temperature estimate in this case becomes
\begin{align}
T_\text{th} =  \frac{\pi^2}{6} \frac{T^2}{\varepsilon_F^*} \left[ 1 - \frac{3}{2}\frac{\rho}{m_n^*} \frac{\partial m_n^*}{ \partial \rho} \right]\ .
\label{eq:thermal_degenerate_int}
\end{align}
Compared to the non-interacting estimate of Eq.~\eqref{eq:thermal_degenerate}, we find a renormalized Fermi energy and an explicit correction due to the density dependence of $m^*_n$.
For pure neutron matter, $m^*_n$ typically differs from the free mass by less than $10 \%$ and the density dependence is relatively mild~\cite{carbone2019}. This explains the relatively small difference between the  two panels of Fig.~\ref{fig:Appendix_thermal}.

The equation for $T_\text{th}$ can be further simplified if an analytical model for the density dependence is employed. In the case of a Skyrme energy density functional, for instance, the density dependence is simple, $m^*_n/m_n = [ 1 + \alpha\, \rho]^{-1}$, with
$\alpha$ a constant~\cite{Chabanat97,Chabanat97un}. Performing the derivative explicitly, we find
\begin{align}
T_\text{th} =  \frac{\pi^2}{6} \frac{T^2}{\varepsilon_F^*} \left[ \frac{5}{2} - \frac{3}{2} \frac{m_n^*}{m_n} \right]\ .
\label{eq:thermal_degenerate_Sk}
\end{align}
We stress that this equation may not be valid for other mean-field models, with more complex density dependence. We show the result of Eq.~\eqref{eq:thermal_degenerate_Sk} with dashed lines in the bottom panel of Fig.~\ref{fig:Appendix_thermal}. As expected, this equation is close to the numerical results in the degenerate regime, above $10^{14}$ ($3 \times 10^{14}$) $\rm g/cm^{3}$  at $T=10$ ($40$) MeV. We have checked that the behavior for other Skyrme interactions is relatively similar, which we take as an indication of the robustness of our arguments.

A possible interpretation of the results in Fig.~\ref{fig:Appendix_thermal} is as follows. Suppose that a hydrodynamics simulation provides a prediction of $T_\text{th}=20$ MeV at $\rho=10^{14}$ g/cm$^3$. If the underlying microphysics model was that of noninteracting pure neutron matter  at that same density, this yields the corresponding horizontal dash-dotted horintal line in the top panel of Fig.~\ref{fig:Appendix_thermal}. The real temperature of a model close to $T_\text{th}=20$ MeV would instead be closer to $30$ MeV at this density, since the solid $T=30$ MeV line crosses the dashed-dotted line corresponding to $T_\text{th}=20$ MeV close to $\rho=10^{14}$ g/cm$^3$. At a more extreme density of  $\rho=10^{15}$ g/cm$^3$, the same analysis at $T_\text{th}=20$ MeV would instead yield a temperature larger than $40$ MeV. In this interpretation, $T_\text{th}$ always underestimates the temperature of the system.

All in all, this exploratory study shows that the use of $T_\text{th}$ from Eq.~\eqref{eq:thermal_temperature} as a proxy for temperature in the hybrid approach should be taken with care. On the one hand, this may be a valid approximation in the low-density, non-interacting regime. On the other, the moment that degeneracy effects set in around saturation density, this equation does not provide a faithful representation of the temperature in the fluid. Our analysis indicates that it provides an underestimation of the real temperature of the system. The results of Fig.~\ref{fig:bulktemp} qualitatively agree with this interpretation. 

\bibliography{refclean}

\end{document}